\documentclass[preprint]{aastex}
\usepackage{natbib}
\shorttitle{Abundances on the RGB of $\omega$ Centauri}
\shortauthors{Stanford et al.}
\newcommand{\wcen}{$\omega$ Cen}
\newcommand{\feh}{[Fe/H]}

\newcommand{\tuc}{47~Tuc}

\begin{document}

\title{Abundances of C, N, Sr and Ba on the red giant branch of $\omega$ Centauri\footnote{This paper includes data gathered with the 6.5 meter Magellan Telescopes located at Las Campanas Observatory, Chile.} }


\author{Laura M. Stanford \altaffilmark{1,2}}
\email{laura@astro.as.utexas.edu}

\and

\author{G. S. Da Costa and John E. Norris \altaffilmark{1}}
\email{gdc, jen@mso.anu.edu.au}

\altaffiltext{1}{Research School of Astronomy and Astrophysics, Australian
  National University, Weston, ACT, 2611, Australia}
\altaffiltext{2}{Current address: McDonald Observatory, The University of Texas at Austin, 1 University Station, C1400, Austin, TX 78712-0259, USA}

\begin{abstract}
Abundances relative to iron for carbon, nitrogen, strontium and barium are
presented for 33 stars on the red giant branch of the globular cluster
$\omega$ Centauri.  They are based on intermediate-resolution spectroscopic 
data covering the blue spectral region analyzed using spectrum synthesis
techniques.  The data reveal the existence of a broad range in the
abundances of these elements, and a comparison with similar data for main
sequence stars enables insight into the evolutionary history of the 
cluster.

The majority of the red giant branch stars were found to be depleted
in carbon, i.e.\ ${\rm [C/Fe]} < 0$, while ${\rm [N/Fe]}$ for the same
stars shows a range of $\sim$1 dex, from ${\rm [N/Fe]} \approx 0.7$ to 1.7
dex.  The strontium-to-iron abundance ratios varied from solar to
mildly enhanced ($0.0 \leq {\rm [Sr/Fe]} \leq 0.8$), with [Ba/Fe] generally
equal to or greater than [Sr/Fe].  The carbon and nitrogen abundance
ratios for the one known CH star in the sample, ROA 279, are
${\rm [C/Fe]}=0.6$ and ${\rm [N/Fe]}=0.5$ dex.

Evidence for evolutionary mixing on the red giant branch is found from
the fact that the relative carbon abundances on the main sequence are
generally higher than those on the red giant branch.  However,
comparison of the red giant branch and main sequence samples shows
that the upper level of nitrogen enhancement is similar in both sets
at ${\rm[N/Fe]} \approx 2.0$ dex.  This is most likely the result of
primordial rather than evolutionary mixing processes.

One red giant branch star, ROA~276, was found to have Sr and Ba
abundance ratios similar to the anomalous Sr-rich main sequence star
S2015448.  High resolution spectra of ROA~276 were obtained with the
Magellan Telescope/{\sc mike} spectrograph combination to confirm this
result, revealing that ROA~276 is indeed an unusual star.  For this
star calculations of the depletion effect, the potential change in
surface abundance that results from the increased depth of the
convective envelope as a star moves from the main sequence to the red
giant branch, strongly suggest that the observed Sr enhancement in
ROA~276 is of primordial origin, rather than originating from a
surface accretion event.

\end{abstract}


\section{Introduction} \label{intro}
\defcitealias{nfm96}{NFM96}
\defcitealias{nd95b}{ND95}
\defcitealias{pan03}{PAN03}
\defcitealias{bw93}{BW93}
\defcitealias{sta06a}{STA06a}
\defcitealias{sta06b}{STA06b}

The unusual variation in abundances of stars within the globular
cluster $\omega$ Centauri has been well studied over the last four
decades.  From the original photographic and photoelectric work of
\citet{woo66} and \citet{cs73} respectively, and the more recent
analysis of CCD photometry by \citet{sol05},  the spread in color
on the red giant branch is apparent with distinct branches
visible in the color-magnitude diagram (CMD).  The large range in metallicity
of over 1 dex has been studied extensively on the red giant branch
(RGB) (\citealt[hereafter NFM96]{nfm96}; \citealt{sk96,lee99, pan00,
rey04, sol05, joh08, cal09}), and to a limited extent on the main sequence and
turnoff (MSTO) area \citep{pio05, sol05, sta06a, kay06, vil07}.

Large ranges in abundance for all elements studied in the cluster have
been found on the RGB (\citealt[hereafter ND95]{nd95b};
\citealt{scl95, smi00, joh08, cal09}).  This contrasts with what is
found in normal globular clusters where typically small ranges for the
light elements (C, N, O, Mg, Al and Na) are seen.  The light elements
in $\omega$ Cen stars show large variations for a given {\feh}
(\citealt{per80}, \citealt{bw93}, hereafter BW93; \citetalias{nd95b}).
The $\alpha$ elemental (Mg, Si, Ca and Ti) abundance ratios with
respect to iron are largely constant for ${\rm [Fe/H]} < -1.0$
(\citetalias{bw93}, \citealt{scl95}; \citetalias{nd95b};
\citealt{smi00}) but [$\alpha$/Fe] then decreases at metallicities
greater than -1.0 \citep{pan02}.  Sodium ([Na/Fe]) and aluminum
([Al/Fe]) abundances are correlated, and both are anticorrelated with
oxygen ([O/Fe]) (\citetalias{bw93}; \citetalias{nd95b};
\citealt{nd95b, smi00}). For lower metallicities (below ${\rm [Fe/H]}
< -0.8$) \citet{smi00} and \citet{cun02} report constant [Cu/Fe].
Above a metallicity of ${\rm [Fe/H]} = -1.2$, however, \citet{pan02}
find an increase in [Cu/Fe] as the metallicity increases.  Variations
in abundance for the neutron-capture elements have also been found.
The s-process element abundance ratios, [s/Fe], increase with
increasing {\feh}, but then are constant above metallicities greater
than $-1.2$ (\citetalias{nd95b}, \citealt{scl95}).  The {\it source}
of these abundance variations can be attributed to several different
possibilities.  The decrease in $\alpha$-elemental abundances at
higher metallicities (${\rm [Fe/H]} \geq -1.0$) can be attributed to
Type Ia supernovae enrichment.  The origin of the s-process enrichment
is possibly due to ejecta from asymptotic giant branch (AGB) stars,
and the iron peak and metal abundance patterns are consistent with
primordial enrichment from Type II supernovae.

At least three different {\it processes} are likely to be involved.
The first are the processes that mix material within the stars
themselves.  These include the first dredge up on the RGB in which the
surface layers of the star are mixed with lower processed material
\citep{ibe65}.  This can, in principle, account for some of the
variations seen within the RGB stars of C, N and O.  These processes,
however, are unlikely to account for the variations seen in elements
heavier than oxygen.

Secondly, an enriched object may have accreted matter onto its surface
layers from either stellar winds from AGB stars or interstellar
material.  As the accretion events only affect the surface layers of
the star, once it ascends the RGB and the convective envelope deepens,
the accreted matter will be mixed with the internal layers, thereby
reducing the level of enrichment.  A comparison of MS and RGB stars
may then show higher abundances for the unevolved MS stars compared to
those on the RGB. \citet{joh09} suggest large binary fractions, and
hence accretion of enhanced material, as the source of the 25\% of
their sample with ${\rm [La/Eu]} \geq 1$.  \citet{cal09} also suggest
the same scenario to explain the $\sim20\%$ of CN strong RGB stars
identified in their sample. However, this is in contrast to that
observed by \citet{may96} who monitored $\sim310$ giants in $\omega$
Cen over a ten year period and found the binary fraction to be as low
as 3 -- 4\%.

The third possibility is that the variations have a primordial origin;
i.e. the stars formed from gas enriched by a combination of AGB stars,
massive stars and/or supernovae.  It is likely that more than one of
these scenarios is at work.

Nevertheless, a comparison between abundances obtained from MS and RGB
stars may help to distinguish among the various processes involved in
the enrichment of this unusual star cluster.  In particular, in the
absence of thermohaline mixing (discussed further in \S5), it can
distinguish whether enrichment of s-process and CNO elements is due to
surface contamination (which would be diluted by the growing
convective envelope as the stars move on to the giant branch) or is
instead uniform throughout the stars. An abundance analysis of MSTO
stars has already been performed by the present authors \citep{sta07}
for carbon, nitrogen, and strontium. In the present paper we
investigate abundances on the RGB in a similar manner.  The sample of
RGB stars and their observations is described \S\ref{rgbs}.  In
\S\ref{sparam} we discuss the stellar parameters and the techniques
used to obtain abundances for C, N, Sr and Ba in the stars.  The
results are presented in \S\ref{abund}.  In \S\ref{disc} we present
comparisons between the RGB and MSTO and discuss the consequences.

\section{The Red Giant Branch Sample and Observations}\label{rgbs}

In order to compare abundances found on the main sequence
\citep{sta07} with those on the RGB, spectra were obtained for a
sample of RGB stars. These objects have $-2.6 < M_{\rm v} < -0.28$,
and are among the brightest RGB stars in the cluster.  All lie above
the luminosity where the first dredge-up occurs and most, if not all,
have luminosities that place them above the bump in the RGB luminosity
function. The objects selected for observation covered a range in
metallicities and several stars were included due to the known nature
of their CH feature, or high s-process abundances.  These objects are
more likely to have higher relative abundances and to be useful for
comparisons with the unusual stars on the MS.  While most of these
objects are ROA objects (Woolley et al. 1966), 11 come from the list
of \citet{pan02}.

The spectra of the RGB stars were obtained on the ANU's 2.3m telescope
at Siding Spring Observatory.  Data were obtained in February 2002,
June 2002, and May 2005 with the Double Beam Spectrograph using 600
and 1200 line/mm gratings.  The stars observed are listed in Table
\ref{tbl1}.  Spectra for eighteen RGB stars were obtained with the 600
line/mm grating.  The 1200 I grating was used to observe 39 RGB stars,
8 of which were in common with those observed with the lower
resolution grating. The low resolution (R600) data covered a wavelength
range 3500--5400{\AA} and had a resolution of 2.2{\AA}
\textsc{fwhm}. The high resolution (R1200) spectra covered a wavelength
range of 3700--4600{\AA} and their resolution was 1.2{\AA}
\textsc{fwhm}.  Reduction and wavelength calibration was performed
using standard routines in IRAF and Figaro.

In addition to the above medium-resolution spectra, high-resolution
data were also obtained for one outstanding object in the sample, ROA
276, together with two other comparison objects, ROA 46
and ROA 150. These spectra were obtained with the Magellan
Telescope/{\sc mike} spectrograph combination during 2007 June 22-23
using a 0.5 arcsec slit, yielding resolving power 47,000, and
resolution 0.12 {\AA} FWHM. This material will be discussed in
\S4.4.3.

\section{Stellar Parameters} \label{sparam} 

Stellar parameters for our sample were obtained primarily from the
literature.  \citetalias{nd95b} listed temperatures ($T_{\rm eff}$),
gravities (${\rm log}g$), metallicities ([Fe/H]) and microturbulent
velocities ($v_{\rm t}$) for a large fraction of the RGB objects for
which we have data, and these were used for consistency between
objects. Temperatures and gravities were also obtained from
\citet{per80} for objects that were not in \citetalias{nd95b}. Stellar
parameters for several RGB stars were also obtained from Pancino
(2003, hereafter PAN03) which are identified by the ``OC'' prefix.
For objects that did not have metallicities listed in
\citetalias{nd95b}, {\feh} was determined using [Ca/H] from
\citetalias{nfm96} where ${\rm [Fe/H]} = {\rm [Ca/H]} - {\rm
  [Ca/Fe]}$. Following \citet{scl95} and \citet{pan00} we adopt the
following [Ca/Fe] relation:

\begin{equation} 
\mbox{[Ca/Fe]} = \left\{ \begin{array}{ll}
0.3 & \mbox{for [Fe/H]} \leq -1.0 \\ 
-0.3\times\mbox{[Fe/H]} & \mbox{for} -1.0<\mbox{[Fe/H]}\leq0.0
\end{array} \right. 
\end{equation}

A small number of objects in our sample had no stellar parameter
information from the sources listed above, or from others in the
literature.  In these cases a metallicity was determined from [Ca/H],
of \citetalias{nfm96}, as described above.  This metallicity was then
used to find other RGB stars in our sample with similar values
($\Delta{\rm [Fe/H]} = \pm0.2$) for which we had stellar parameters.  A
least squares fit was then performed in V--${\rm log}g$ space for the chosen
comparison stars to obtain a gravity for the star from its V
magnitude.  Similarly, a temperature was obtained from a least squares
fit in $(B-V) - T_{\rm eff}$ space.  As most RGB stars have a
microturbulent velocity of $\sim2.0~{\rm kms}^{-1}$, this value was
assumed for these objects.  However, there remained a small number of
stars for which we were unable to obtain stellar parameters (ROA 55,
577; and OC 230189, 242056, 250606, 312058, 801639) and consequently
these stars were not included in the analysis.

Table \ref{tbl2} lists the stellar parameters used for each star, with
ROA or OC number in column 1, V magnitude and B--V color in columns 2
and 3, and $T_{\rm eff}$, ${\rm log}g$, metallicity and microturbulence in
columns 4--7, respectively. The sources of the stellar parameters are
listed against each star for clarification in column 8.

Abundances of [O/Fe] and [Ca/Fe] were obtained from \citetalias{nd95b}
and \citetalias{pan03}.  In the instances where these were not available,
the  abundance ratios of these elements followed equation 1 above.

Two objects, ROA 336 and 517, observed with the higher resolution
grating, had high metallicities (${\rm [Ca/H]} = -0.33$ and 0.02 respectively,
\citealt{nfm96}).  Spectrum synthesis was attempted but it was found
that for these cool, metal-rich stars the models do not produce an
adequate fit to the observed spectra. Consequently, these objects were
not analyzed further. The remaining sample then consisted of 23
objects with high resolution (1.2\AA) and 16 stars with low resolution
(2.2\AA) spectra, with 7 stars in common.

\section{Abundances} \label{abund}

The abundance analysis used spectrum synthesis techniques for each of
the objects, adopting the stellar parameters listed in Table
\ref{tbl2}.  \citet{kur93} stellar models were employed with atomic
line lists from Bell (2000, private communication) and Kurucz
molecular lines lists.  As with our study of abundances on the main
sequence/subgiant branch \citep{sta07} the synthesis code developed by
Cottrell \citep{cn78} was used to generate synthetic spectra.  These
were broadened to 1.1{\AA} and 2.2{\AA} to match the observed higher
and lower resolution data, respectively.  The solar abundances and gf
values for CH, CN, Sr \textsc{ii} 4077{\AA}, Sr \textsc{ii} 4215{\AA}
and Ba \textsc{ii} 4554{\AA} used in \citet{sta07} were again
employed.

As for the MSTO data, the CH at 4300{\AA} feature was analyzed first.
The resulting [C/Fe] was then adopted in the analysis of the CN
features.  Due to the cooler nature of the RGB stars in comparison
with the MSTO stars, the violet CN at 3883{\AA} is often saturated
making any abundance determination from this band uncertain.  Therefore
the blue CN band at 4216{\AA} was used for the RGB stars.  The effect
this has on the derived abundances is minimal, and was tested using an
RGB object (ROA 40) with low N enhancement (${\rm [N/Fe]} = 0.6$).  For this
star, the violet CN band was not saturated and consistent results were
found for [N/Fe] from both bands.

Lastly, the Sr \textsc{ii} 4077{\AA}, Sr \textsc{ii} 4215{\AA} and Ba
{\sc ii} 4554{\AA} features were analyzed.  In contrast to the MSTO
data, Ba abundances were able to be determined for each RGB star.
This was due to the higher resolution and S/N of the data and/or
cooler temperatures.  The Sr \textsc{ii} 4215{\AA} line was not
analyzed if ${\rm [N/Fe]} > 0.5$ owing to contamination from the 4215{\AA} CN
features.  In the synthetic spectra, several lines (e.g. iron,
titanium, samarium, CH) were present near the Sr \textsc{ii} 4077{\AA}
feature ($\lambda\lambda4075-4077{\rm \AA}$) that were more clearly
separated in the observed high resolution (R1200) spectra.  In the
lower resolution data (R600) these features were blended with the Sr
\textsc{ii} line making the abundance determined from the R600 data
slightly more uncertain.  The Sr abundances were determined from the
models both with and without these lines, which led to the conclusion
that they did not significantly affect the [Sr/Fe] measurements. This
was particularly the case when the Sr feature was strong.

The effect of hyperfine splitting (hfs) on the derived abundance from
the Ba \textsc{ii} 4554{\AA} line was investigated following
\citet{nrb97}.  Two line lists were used: one that included hfs and
isotopic components for Ba \textsc{ii} 4554{\AA}, and one that did
not.  A comparison of abundances was made for a series of stellar
parameters and [Ba/Fe] spanning the same range as for the RGB sample.
Little difference ($\Delta{\rm [Ba/Fe]} < 0.05$) was found between the
abundances obtained from the spectra with and without the inclusion of
isotopes and hyperfine splitting.

\subsection{Error Analysis}

An error analysis was performed by varying the temperature, gravity,
metallicity and microturbulence, and redetermining the abundances for
several stars in both the R600 and R1200 samples.  The
parameters were varied by $\Delta T = \pm100{\rm K}$,
$\Delta{\rm log}g = \pm0.2$, $\Delta{\rm [Fe/H]} = \pm0.2$ and
$\Delta{\rm v}_{\rm t} = \pm0.3$.

The derived carbon abundance is dependent on the O abundance, which
was taken from \citealt{nd95} and \citealt{pan03} when available.  The
oxygen abundance [O/Fe] was varied by $\pm0.2$ and the spectrum
synthesis was repeated to determine the error in [C/Fe] due to the
possible uncertainty in the oxygen abundance. This error was then
included in the final error of [C/Fe].  These individual errors were
added in quadrature and gave a final error $\Delta{\rm [C/Fe]}_{\rm
  R600} = 0.26$ for the R600 data and $\Delta{\rm [C/Fe]}_{{\rm
    R1200}} = 0.22$ for the R1200.  As the nitrogen abundance
determination was dependent on the adopted [C/Fe], the final [N/Fe]
error also included the uncertainty in [C/Fe].  The errors in the N
abundance were $\Delta{\rm [N/Fe]}_{{\rm R600}} = 0.35$ and
$\Delta{\rm [N/Fe]}_{{\rm R1200}} = 0.31$ for the R600 and R1200 data,
respectively.  The uncertainties in Sr and Ba abundances were
determined to be $\Delta{\rm [Sr/Fe]}_{{\rm R600}} = 0.37$ and
$\Delta{\rm [Sr/Fe]}_{{\rm R1200}} = 0.35$, and $\Delta{\rm
  [Ba/Fe]}_{{\rm R600}} = 0.31$ and $\Delta{\rm [Ba/Fe]}_{{\rm R1200}}
= 0.28$.

In addition to the stellar parameters and oxygen abundance errors, a
measurement uncertainty exists in the determination of the individual
abundances.  This was estimated by visual inspection of the quality of
the fit for each star for carbon, nitrogen, strontium and barium, and
combined with the general errors above.  These uncertainties are
listed for each star individually in Table \ref{tbl3} for the low
resolution data and Table \ref{tbl4} for the R1200 ones.

\subsection{Abundances of C, N, Sr and Ba}

\begin{figure}
\begin{center}
\includegraphics[width=7cm,angle=0]{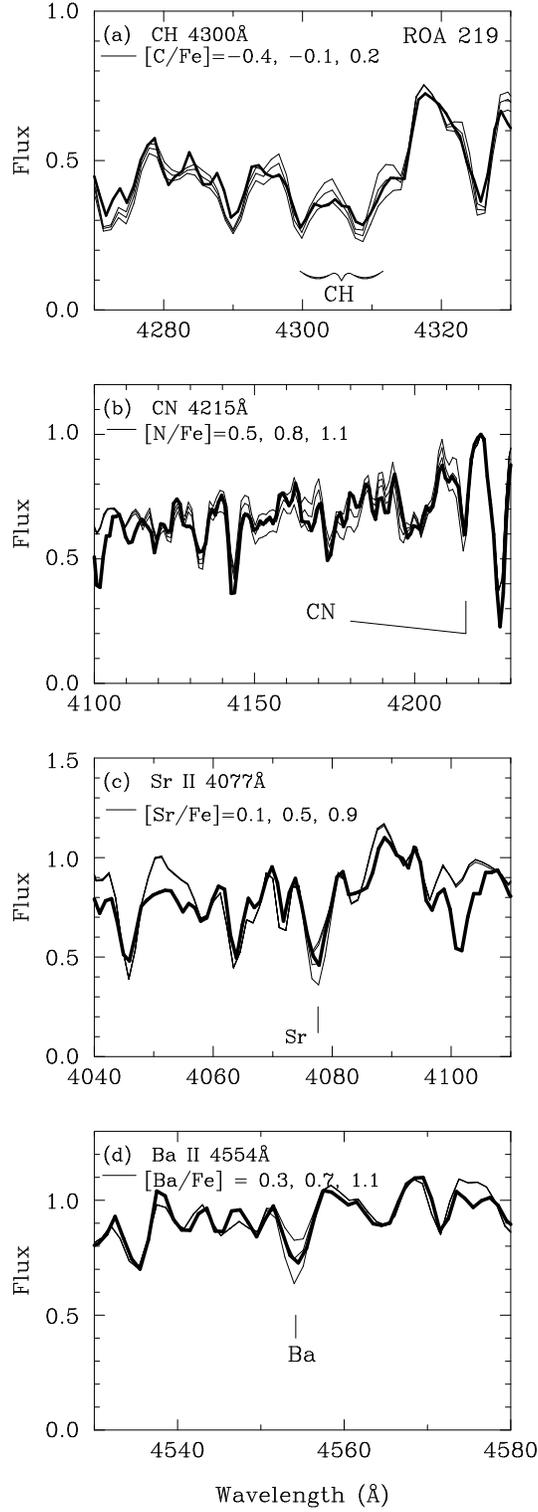}
\caption{ Observed spectrum (black line) of ROA 219 with synthetic
  spectra (grey lines).  It has a resolution of 2.2{\AA}.  The synthetic spectra in panel (a) have ${\rm [C/Fe]} = -0.4, 0.0, 0.2$. The synthetic spectra in panel (b) adopt ${\rm [N/Fe]} = 0.5, 0.8, 1.1$, with ${\rm [C/Fe]} = -0.1$. Panel (c) and (d) shows synthetic spectra with ${\rm [Sr/Fe]} = 0.1, 0.5, 0.9$, and ${\rm [Ba/Fe]} = 0.3, 0.7, 1.1$ respectively. \label{p219}
}
\end{center}
\end{figure}

The low resolution (R600) abundance results are shown in Table
\ref{tbl3}.  For all but one star in this data set the C abundance was
found to be depleted (${\rm [C/Fe]} < 0.0$).  The object with the
highest C abundance was the CO-strong star ROA 179 with ${\rm [C/Fe]}
= 0.1$. Large overabundances in nitrogen were seen in C-depleted
objects, and moderate enhancements of Sr and Ba were found.  In
general, the Ba enhancements were greater or equal to those for Sr,
with ${\rm [Ba/Sr]} \geq 0.0$.  Figure \ref{p219} shows an example of
low resolution spectra of the CO-strong star ROA 219 with the spectrum
synthesis results.

Similar patterns to those found in the R600 data are also seen in the
R1200 data (Table \ref{tbl4}).  Most stars are C depleted and enhanced
in nitrogen. Sr and Ba abundances again show moderate enhancements
with ${\rm [Ba/Sr]} \geq 0.0$ for the majority of stars.

\begin{figure}
\begin{center}
\includegraphics[width=6cm,angle=0]{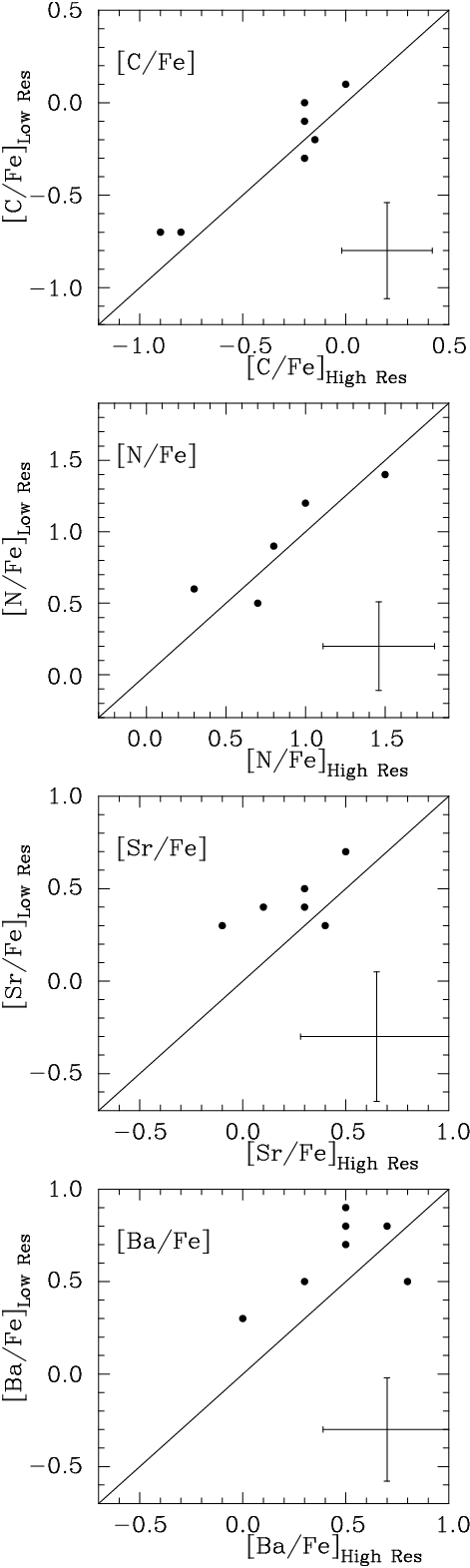}
\caption{Comparison between R600 and R1200 carbon,
nitrogen, strontium and barium abundances for stars in common.
\label{pc612}}
\end{center}
\end{figure}

The six stars for which both R600 and R1200 data were obtained (see
Figure \ref{pc612}) show mean abundance differences (with the standard
error of the mean), $\Delta({\rm R600-R1200}) = 0.08\pm0.05$ for
[C/Fe], $0.10\pm0.06$ for [N/Fe], $0.18\pm0.07$ for [Sr/Fe], and
$0.17\pm0.09$ for [Ba/Fe].  We will need to keep in mind the
possibility of a systematic offset between the R600 and R1200 [Sr/Fe]
values in the subsequent discussion.

\subsection{Comparisons With Other Studies} \label{compos}

The abundances determined here can be compared with those in the
literature for the same stars.  \citetalias{nd95b} determined [C/Fe]
from the G band, [N/Fe] from the blue CN band and [Ba/Fe] from three
lines (not including the one used here) for a large fraction of our
sample. Although they did not observe Sr in their investigation of RGB
stars, other light s-process elements such as yttrium and zirconium
were included and these can be compared with the Sr abundances
determined here for stars in common between the two studies.
\citetalias{bw93} measured carbon (from the G band) and nitrogen (from the
red CN band at 7950{\AA}), among other elements, and have four stars
in common (ROA 48, 74, 84, 213) with this study.  The RGB stars with
OC designations come from \citetalias{pan03}, and the results found here
can be compared with the abundances of Ba, Y and Zr published there.

\begin{figure}
\begin{center}
\includegraphics[width=6.0cm,angle=0]{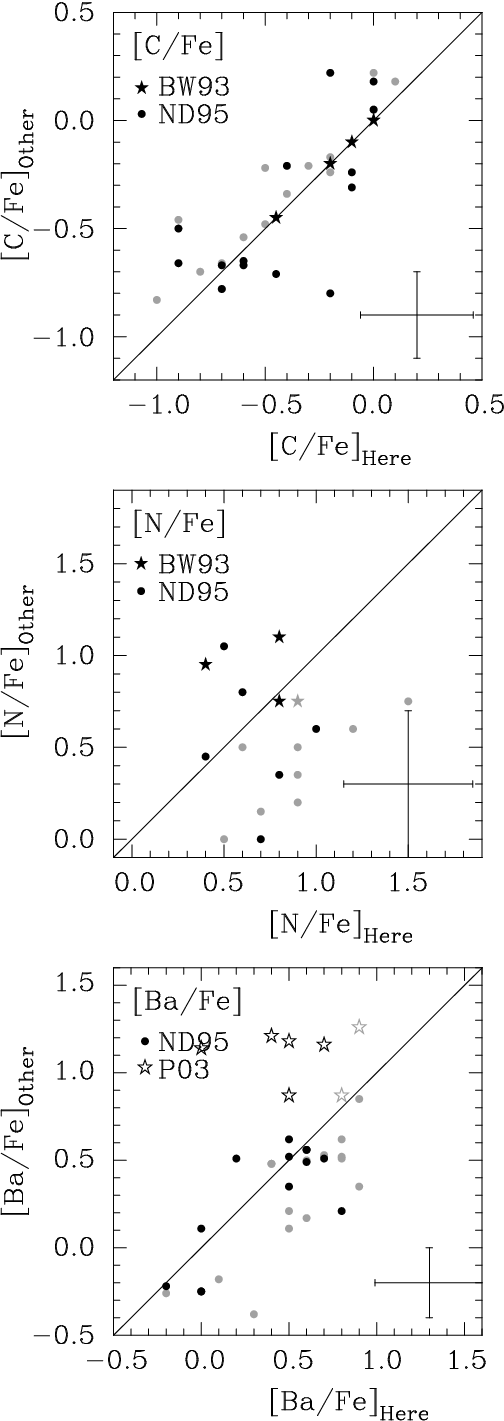}
\caption{ Carbon, nitrogen and barium abundance comparisons
  between the R600 work (grey) and R1200 (black)
  on one hand, and \citetalias{nd95b},\citetalias{bw93} and
  \citetalias{pan03} on the other. The filled dots represent the
  comparisons between this work and \citetalias{nd95b}, the filled
  stars for \citetalias{bw93} and the open stars for
  \citetalias{pan03}.  $\Delta{\rm [X/Fe]} = {\rm [X/Fe]}_{\rm This study} - {\rm [X/Fe]}_{\rm Other studies}$
\label{pc600} }
\end{center}
\end{figure}

Figure \ref{pc600} shows the comparisons between the present study on
the one hand and \citetalias{nd95b} and \citetalias{bw93}, on the
other, for [C/Fe] and [N/Fe].  The barium abundances were compared
with those found in \citetalias{nd95b} and \citetalias{pan03} for
stars in common. The R600 data show good agreement for the carbon
abundances with \citetalias{nd95b} and \citetalias{bw93}.  The mean
abundance differences, in the sense this study$-$other study, were
$\langle\Delta{\rm [C/Fe]}\rangle_{{\rm ND}} = -0.11\pm0.13$ and
$\langle\Delta{\rm [C/Fe]}\rangle_{{\rm BW}} = -0.10$ for the one star
in common.  The R1200 data show a similar pattern where
$\langle\Delta{\rm [C/Fe]}\rangle_{{\rm ND}} = -0.11\pm0.17$ and
$\langle\Delta{\rm [C/Fe]}\rangle_{{\rm BW}} = -0.19\pm0.19$.  The
nitrogen abundances determined from both the R600 and the bulk of the
R1200 data show offsets to the \citetalias{nd95b} study of
$\langle\Delta{\rm [N/Fe]}\rangle_{{\rm ND}} = 0.52\pm0.20$ and
$\langle\Delta{\rm [N/Fe]}\rangle_{{\rm ND}} = 0.12\pm0.19$,
respectively.  Similar results are found when we compare our
abundances with those of \citetalias{bw93} with $\langle\Delta{\rm
  [N/Fe]}\rangle_{{\rm BW}} = 0.15$ for the one star in common in the
R600 sample, and $-0.27\pm0.19$ for the R1200 data, respectively.  The
reason for the offset of $\sim0.5$ dex in [N/Fe] for the R600 data is
probably due to the lower resolution of the data.  As for [Sr/Fe], we
need to keep in mind the possibility of a systematic offset between
the R600 and R1200 [N/Fe] values in the subsequent discussions.

The barium abundances obtained from the R600 data are greater by
$\langle\Delta{\rm [Ba/Fe]}\rangle_{\rm ND} = 0.24\pm0.21$ and
$\langle\Delta{\rm [Ba/Fe]}\rangle_{\rm P03} = -0.22\pm0.21$ compared
with other studies.  This is most likely due to the low resolution of
our data and the blended nature of the Ba \textsc{ii} 4554{\AA}
feature used for the analysis.  The R1200 data show better agreement
with \citetalias{nd95b} with $\langle\Delta{\rm [Ba/Fe]}\rangle_{\rm
  ND} = 0.08\pm0.22$.  The determination of [Ba/Fe] is somewhat easier
for these spectra due to the reduced blending of the feature.  From
the four stars in common with \citetalias{pan03} the barium abundances
determined here for the R1200 data are lower by $\langle\Delta{\rm
  [Ba/Fe]}\rangle_{ \rm P03} = -0.69\pm0.18$.  As the same stellar
parameters were used in both studies, the difference in abundance is
most likely due to the differing nature of the analysis.  Spectrum
synthesis techniques were employed here for the Ba \textsc{ii}
4554{\AA} line, while \citetalias{pan03} used an equivalent width
analysis and different lines.

\begin{figure}
\begin{center}
\includegraphics[width=6.0cm,angle=0]{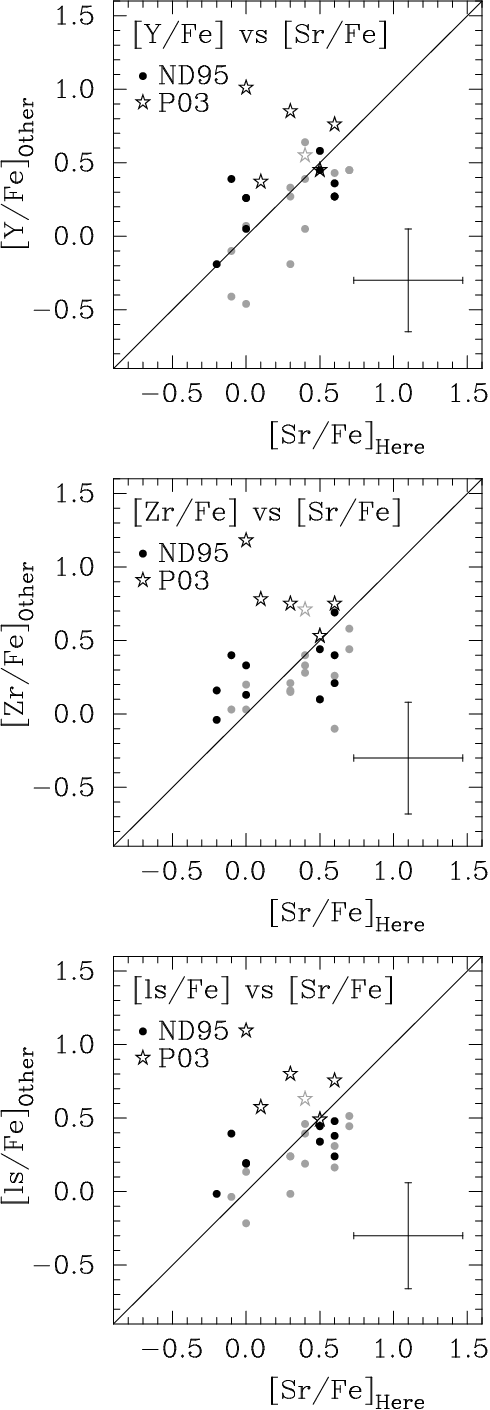}
\caption{ Light s-process element (Y, Zr) comparison between the R600
  work (grey dots) and R1200 work (black dots) on one hand, and
  \citetalias{nd95b} (filled dots) and \citetalias{pan03} (open stars)
  on the other. [ls/Fe] is an average of abundances of [Zr/Fe] and
  [Y/Fe].  $\Delta{\rm [X/Fe]} = {\rm [X/Fe]}_{\rm This study} - {\rm [X/Fe]}_{\rm Other studies}$ . Please note the filled star in the top panel is actually a unfilled star and a filled dot at the same
point.
  \label{pc600b} }
\end{center}
\end{figure}

Figure \ref{pc600b} shows the comparison between the Sr abundances
obtained here compared with [Y/Fe] and [Zr/Fe] obtained by
\citetalias{nd95b} and \citetalias{pan03}.  The lower panels show the
averaged [Y/Fe] and [Zr/Fe] abundances designated here as [ls/Fe].
The mean differences between the R600 data$-$other studies are
$\langle\Delta{\rm [ls/Fe]}\rangle_{\rm ND} = 0.14\pm0.17$ and
$\langle\Delta{\rm [ls/Fe]}\rangle_{\rm P03} = -0.11\pm0.17$.  The
R1200 data show similar abundance patterns for Sr compared with the
other light elements studied in \citetalias{nd95b}, with
$\langle\Delta{\rm [ls/Fe]}\rangle_{\rm ND} = 0.20\pm0.30$.  The
comparisons of the light s-process elements with \citetalias{pan03}
show $\langle\Delta{\rm [ls/Fe]}\rangle_{\rm P03} = -0.32\pm0.25$
similar to that found for the barium abundances.

\subsection{Results} \label{results}

\subsubsection{Carbon and Nitrogen}

\begin{figure}
\begin{center}
\includegraphics[width=7.0cm,angle=0]{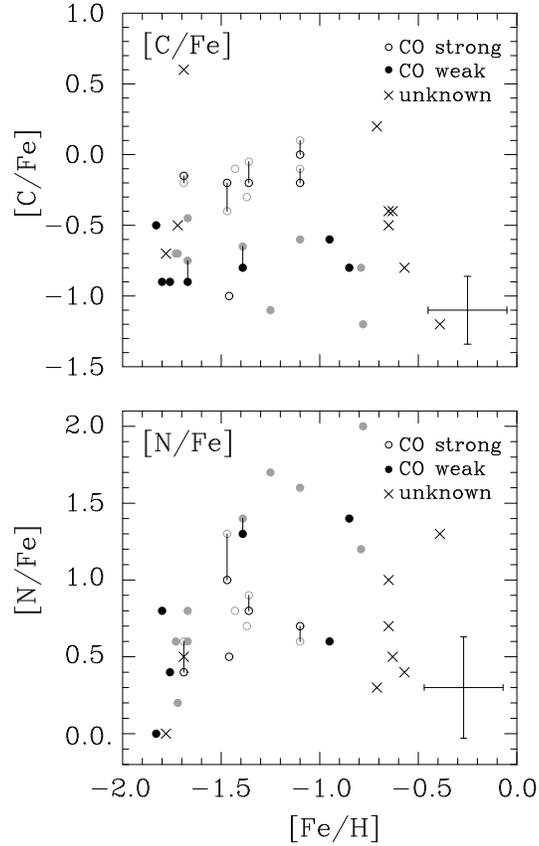}
\caption
{ Carbon and nitrogen abundances plotted as functions of
 metallicity for the R600 (grey dots) and R1200 (black dots)
 resolution data.  The open circles represent the CO-strong stars, and
 closed circles represent the CO-weak stars.  The stars for which
 there is no information regarding the CO nature are represented by
 crosses. Abundances of stars in common between the R600 and R1200
 resolution analysis are joined by solid lines.\label{pcna} }
\end{center}
\end{figure}

The carbon and nitrogen abundances for the R600 (grey points) and R1200
(black points) resolution data are shown as functions of metallicity
in Figure \ref{pcna}. Objects in common between the R600 and R1200
data sets are joined by solid lines.  The abundances have been plotted
according to whether they are CO-strong (open circles) or weak (closed
circles) following the convention of \citetalias{nd95b}.  Crosses are
plotted for which there is no information regarding their CO status.
In general, a CO-strong star is found to have high [C/Fe] and
relatively low [N/Fe]. Conversely, a CO-weak star shows low carbon
abundances and high nitrogen ones.

Figure \ref{pcnb} plots the carbon abundances against those of
nitrogen for the R600 (grey dots) and R1200 (black dots) resolution
data. Again, objects in common between the two data sets are joined by
solid lines.  An anticorrelation can be seen within the data, but
there is no evidence for bimodality.  A least squares fit was
performed on the data, indicated by the solid line. The p-value for
the least squares fit was 0.183, indicating some correlation but only
a very moderate association between the carbon and nitrogen
abundances\footnote{The p-value for the least squares fit indicates
  the probability that the sample was drawn from the population being
  tested given the assumption that the null hypothesis is true.  A
  p-value of 1 gives an 100\% probability of the null hypothesis being
  true, while a ${\rm p-value} = 0.05$ gives a 5\% probability. In this case,
  the null hypothesis was that the [C/Fe] and [N/Fe] abundances were
  unrelated.}.

The RGB star ROA 279 is a known CH star \citepalias{nd95b}.  No
previous carbon or nitrogen abundance determinations could be found in
the literature for this object (\citetalias{nd95b} analyzed lines of
other elements but did not report carbon, nitrogen or oxygen
abundances).  The O abundance for ROA 279 is not known and as this has
an impact on the derived C abundance, an analysis was performed to
determined its effect.  Using oxygen abundances in the range ${\rm
  [O/Fe]} = -0.6 - +0.6$ led to a spread in C of 0.8 dex, with ${\rm
  [C/Fe]} = -0.1 - +0.7$.  The oxygen abundance was assumed to be
${\rm [O/Fe]} = 0.3$, and this led to ${\rm [C/Fe]} = 0.6$ and ${\rm
  [N/Fe]} = 0.5$.  The Sr and Ba abundances determined for ROA 279
were ${\rm [Sr/Fe]} = {\rm [Ba/Fe]} = 0.6$.  \citet{bd74} analyzed two
CH stars, ROA 55 and ROA 70, in {\wcen}, and found ${\rm [C/H]} =
-0.8$, ${\rm [N/H]} = 0.0$ and assumed ${\rm [O/H]} = -0.3$.
Adopting ${\rm [Fe/H]} = -1.9$ for ROA 70 \citep{gra82}, the carbon
and nitrogen abundance ratios are ${\rm [C/Fe]} = 1.0$ and ${\rm
  [N/Fe]} = 1.8$.  While the abundance of C found here (${\rm [C/Fe]}
= 0.6$) for ROA 279 is less than these CH stars analyzed by
\citet{bd74}, it is still considerably higher than in other RGB stars
in {\wcen}, including the CO-strong objects.  The enhancement of the
s-process elements is comparable to that of the other CH stars, ROA 55
and 70 \citep{gra82}, field CH stars and other stars with s-process
enhancement in the cluster.  The nitrogen abundances obtained,
however, are over one dex different.  This is an interesting
discrepancy and warrants further investigation.  The enrichment of C
and s-process material in field CH stars is generally thought to come
from mass transfer in binary systems \citep{mw90}, and may be the case
here as well.

\subsubsection{Strontium and Barium}

\begin{figure}
\begin{center}
\includegraphics[width=7.0cm,angle=0]{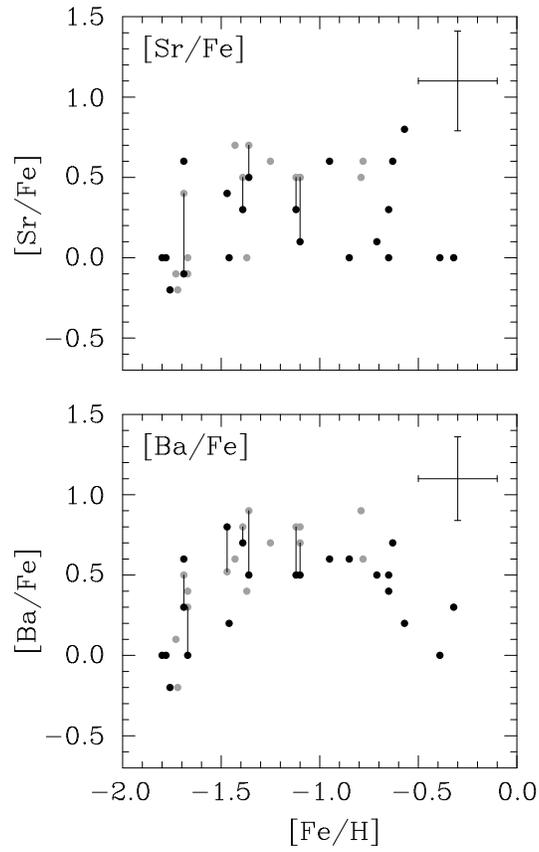}
\caption{ 
Strontium and barium abundances plotted as functions of metallicity
for the R600 (grey) and R1200 (black)  data. Abundances of
stars in common between the R600 and R1200 resolution analysis are
joined by solid lines.
\label{psrbaa} }
\end{center}
\end{figure}

The Sr and Ba abundances are plotted as a function of metallicity in
Figure \ref{psrbaa} for both the R600 (grey) and R1200 (black)
resolution data.  The steep incline and then roughly constant value
seen in the Ba abundance plot is similar to that seen in
\citetalias{nd95b}.  At higher metallicities a downward trend in the
Ba abundances is also seen which is not present in other studies.
These objects are the most metal-rich ones in our sample, and several
come from the studies of \citetalias{pan03}.  In \S \ref{compos} a
comparison was made between the barium abundances determined here and
those from \citetalias{pan03}.  It was found that for stars in common
[Ba/Fe] values determined here were lower than those found in
\citetalias{pan03} ($\langle{\rm [Ba/Fe]}_{\rm R600} \rangle \sim -0.22$ and $\langle{\rm [Ba/Fe]}_{\rm R1200} \rangle \sim -0.69$).
This may account for the downward trend seen in the barium abundances
at higher metallicities.

\begin{figure}
\begin{center}
\includegraphics[width=7.0cm,angle=0]{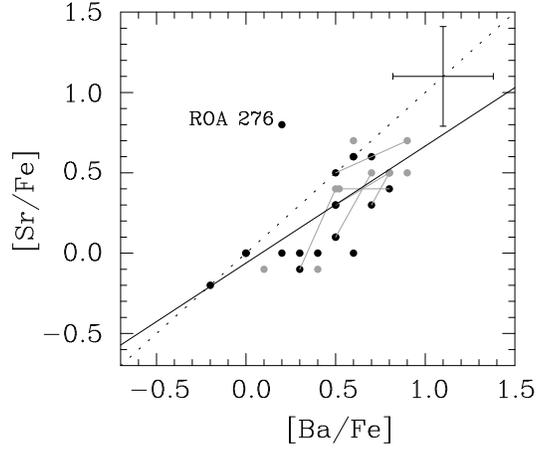}
\caption{ Strontium
abundance plotted as a function of barium abundance for the R600
(grey) and R1200 (black) resolution samples. Abundances of stars in
common between the R600 and R1200 analysis are joined by
solid lines. The solid line is the least squares fit to the data.  The
dotted line is the 1:1 line for reference.
\label{psrbab} }
\end{center}
\end{figure}

Figure \ref{psrbab} plots the strontium abundance against that of
barium for the R600 (grey) and R1200 (black) resolution data.  Objects
in common between the two data sets are joined by a solid line.  This
plot shows a clear correlation between [Sr/Fe] and [Ba/Fe]. A
one-to-one line is drawn for reference as a dotted line.  A least
squares fit was performed on the data, indicated by the solid
line. The p-value for the least squares fit was $<0.001$, indicating a
very high probability that there is a correlation between the
strontium and barium abundances\footnote{ In this case, the null
  hypothesis was that the [Sr/Fe] and [Ba/Fe] abundances were
  unrelated.}.

\begin{figure}
\begin{center}
\includegraphics[width=8.0cm,angle=0]{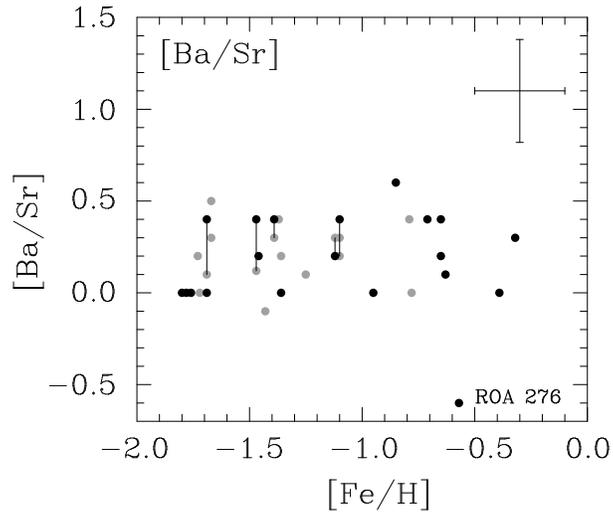}
\caption{ Ratio of heavy to light neutron capture elements plotted as
a function of metallicity for the R600 (grey) and R1200 (black)
samples. Abundances of stars in common between the R600 and R1200
analysis are joined by solid lines.
\label{phsls} }
\end{center}
\end{figure}

The ratio of the heavy s-process elements (e.g. Ba) to light ones
(e.g. Sr) gives an indication of the level of neutron exposure, with a
higher value indicating a higher exposure.  Figure \ref{phsls} plots
the ratio [Ba/Sr], designated here as [hs/ls], as a function of
[Fe/H].  Most objects have ${\rm [hs/ls]} \geq 0.0$, with one
exception, ROA 276, described in more detail in the following
section. This star shows the surprisingly low value ${\rm [Ba/Sr]} =
{\rm [hs/ls]} = -0.6$.

\subsubsection{ROA 276} \label{roa276}

\begin{figure}
\begin{center}
\includegraphics[width=12.0cm,angle=0]{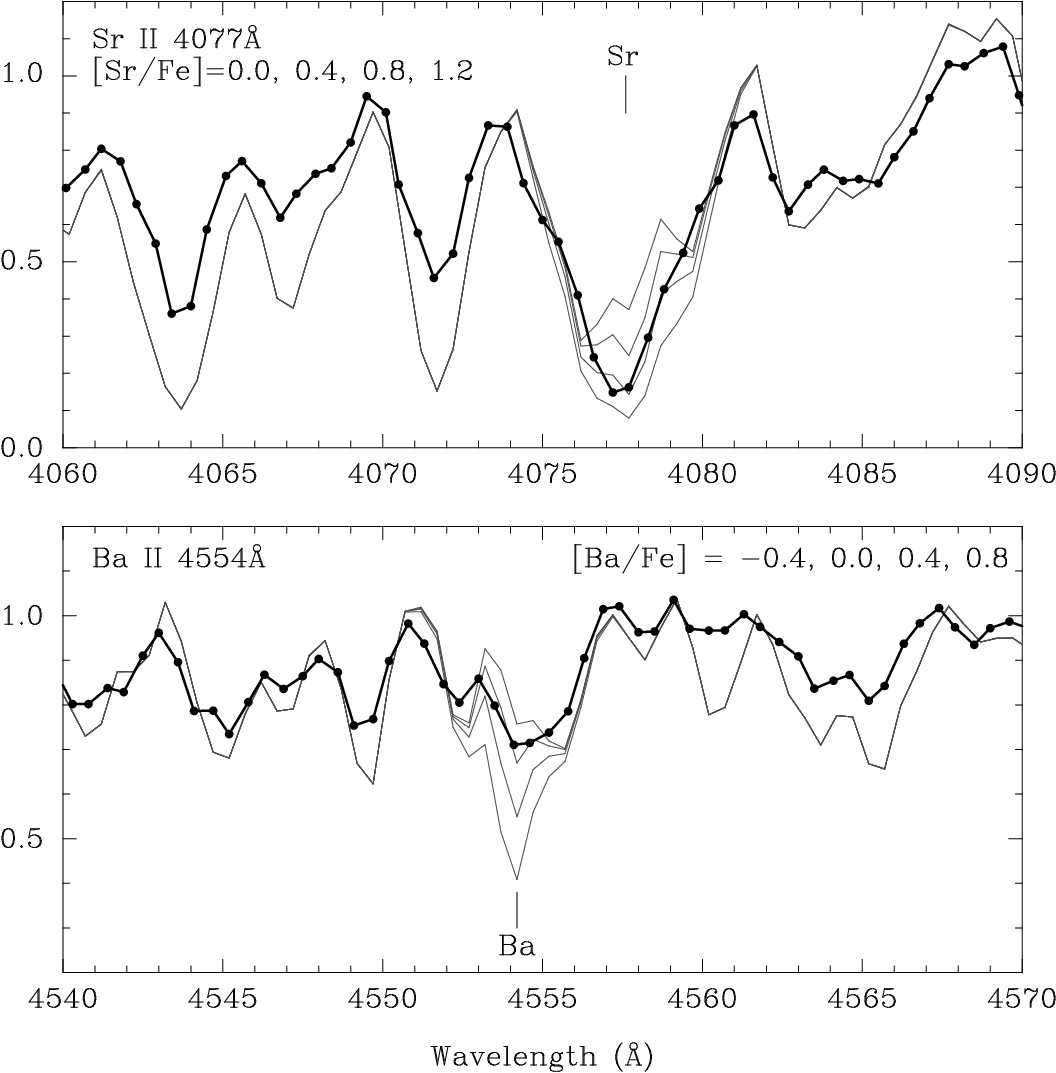}
\caption{ Observed 1200I grating spectrum of ROA 276 (black line) with
  synthetic spectra (grey lines).  The stellar parameters used were  $T_{\rm eff} = 4000$, ${\rm log}g = 0.7$, ${\rm [Fe/H]} = -0.57$,  ${\rm v}_{t} = 2.0$.  \label{p276} }
\end{center}
\end{figure}

While most stars in the sample had measured [Ba/Fe] equal to or
greater than [Sr/Fe], ROA 276 deviates from this general trend with
significantly larger [Sr/Fe] than [Ba/Fe], as seen in Figure
\ref{psrbab}. The reader may recall that a similar [Ba/Sr] ratio was
found before in the $\omega$ Cen main sequence star S2015448,
described in \citet{sta06b}.  ROA 276 is metal rich with ${\rm [Fe/H]}
= -0.57$, has depleted carbon (${\rm [C/Fe]} = -0.80$), and enhanced
nitrogen (${\rm [N/Fe]} = 0.4$).  The observed and synthetic spectra
in the region of Sr {\sc ii} 4077{\AA} and Ba {\sc ii} 4554{\AA} are shown
in Figure \ref{p276}.

The stellar parameters for this object were derived by interpolation
of $T_{\rm eff}$ and ${\rm log}g$ given its metallicity converted from
[Ca/H] \citepalias{nfm96} as described in \S\ref{sparam}.  When
analyzing the Sr \textsc{ii} 4077{\AA} feature in the observed spectra
it was noticed that many of the neighboring lines were too strong,
indicating the metallicity was too high or the temperature too low.
By decreasing the metallicity to ${\rm [Fe/H]} = -1.4$, and
redetermining the temperature to be 4200K (as opposed to 4000K), an
improved fit to these lines was found.  The effect this had on the
abundances was to increase [Sr/Fe] from 0.8 to 1.8 and [Ba/Fe] from
0.2 to 0.6.  Despite the change in stellar parameters the unusual
[Ba/Sr] ratio is still present.  Although the original stellar
parameters and abundances have been recorded in each of Tables
\ref{tbl3} and \ref{tbl4}, it should be kept in mind that the
temperature and metallicity have been estimated and require more
accurate determination.  Further analysis of this object is required
to accurately determine its stellar parameters and to confirm the
s-process abundance pattern found here.

As a consequence of the abundance result obtained with the
intermediate resolution spectra, higher resolution data were obtained
for ROA 276 along with two other $\omega$ Cen RGB stars, ROA 46 and
ROA 150.  These data were obtained with the Magellan (Clay)
telescope/{\sc mike} echelle spectrograph combination (see
http://www.ucolick.org/$\sim$rab/MIKE/usersguide.html) during 2007 Jun
22---23. We used a 0.5'' slit and resolving power of 47 000.  The
resolution of the data are $~$0.12{\AA} FWHM. Here we concentrate on
the wavelength regions 4065--4085{\AA} and 4545--4565{\AA} to cover
the Sr {\sc ii} 4077{\AA} line and the Ba {\sc ii} 4554{\AA} one.
Hyperfine splitting for the Ba {\sc ii} 4554{\AA} line was taken into
account and followed the s-process abundance pattern from
\citet{arl99}.

\begin{figure}
\begin{center}
\includegraphics[width=12.0cm,angle=0]{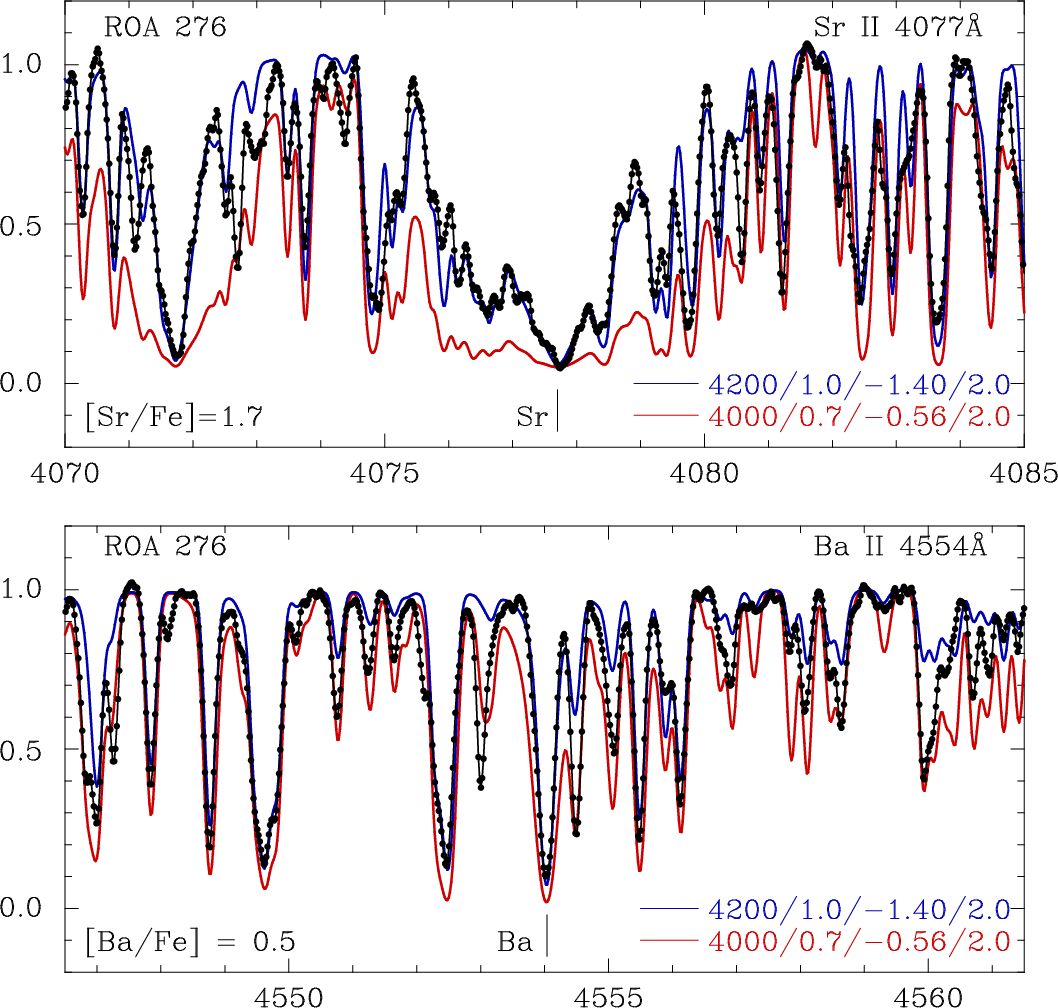}
\caption{ High resolution observed {\sc mike} spectrum of ROA 276 (black dotted line) plotted with two synthetic spectra with differing stellar parameters. The solid blue line has $T_{\rm eff}$/${\rm log}g/[Fe/H]/{\rm v}_{t}$ = 4200/1.0/--1.40/2.0 and the red line has  $T_{\rm eff}$/${\rm log}g/[Fe/H]/{\rm v}_{t}$ = 4000/0.7/0.56/2.0.
  \label{psc} }
\end{center}
\end{figure}

The synthetic spectra were produced using the models and linelists
employed for the 2.3m spectra as described in \S\ref{abund}.  As
mentioned previously, the stellar parameters are not well known for
ROA 276.  Synthetic spectra produced using the parameters determined
above ($T_{\rm eff} = 4000K$, ${\rm log}g = 0.7$, ${\rm [Fe/H]} =
-0.57$) did not match the observed spectra very well with the lines
produced in the synthetic spectra being deeper than those observed.
To illustrate this point, in Figure \ref{psc} we compare the observed
{\sc mike} high resolution data for both Sr II 4077{\AA} (upper panel)
and Ba II 4554{\AA} (lower panel) with the synthetic spectra having
stellar parameters ($T_{\rm eff} = 4000K$, ${\rm log}g = 0.7$, ${\rm
  [Fe/H]} = -0.57$) and ($T_{\rm eff} = 4200K$, ${\rm log}g = 1.0 $,
 ${\rm [Fe/H]} = -1.4$) in both panels.

To test the robustness of the results based on the stellar parameters,
synthetic spectra were produced with a range in temperature and
metallicity (surface gravity corresponded to the temperature based on
the star's position in the CMD).  Four effective temperatures ($T_{\rm
  eff} = 3800K, 4000K, 4200K, 4400K$), and three metallicities (${\rm
  [Fe/H]} = -0.57 , -1.0, -1.4$) were used.  The [Ba/Sr] ratio ranges
between $-1.1$ and $-1.4$ for each of these models. As shown in Table
\ref{tbl6} [Ba/Sr] is largely independent of the input stellar
parameters.  More accurate stellar parameters will be determined in a
forthcoming paper where the red high resolution {\sc mike} data are
analyzed and results presented.

\begin{figure}
\begin{center}
\includegraphics[width=12.0cm,angle=0]{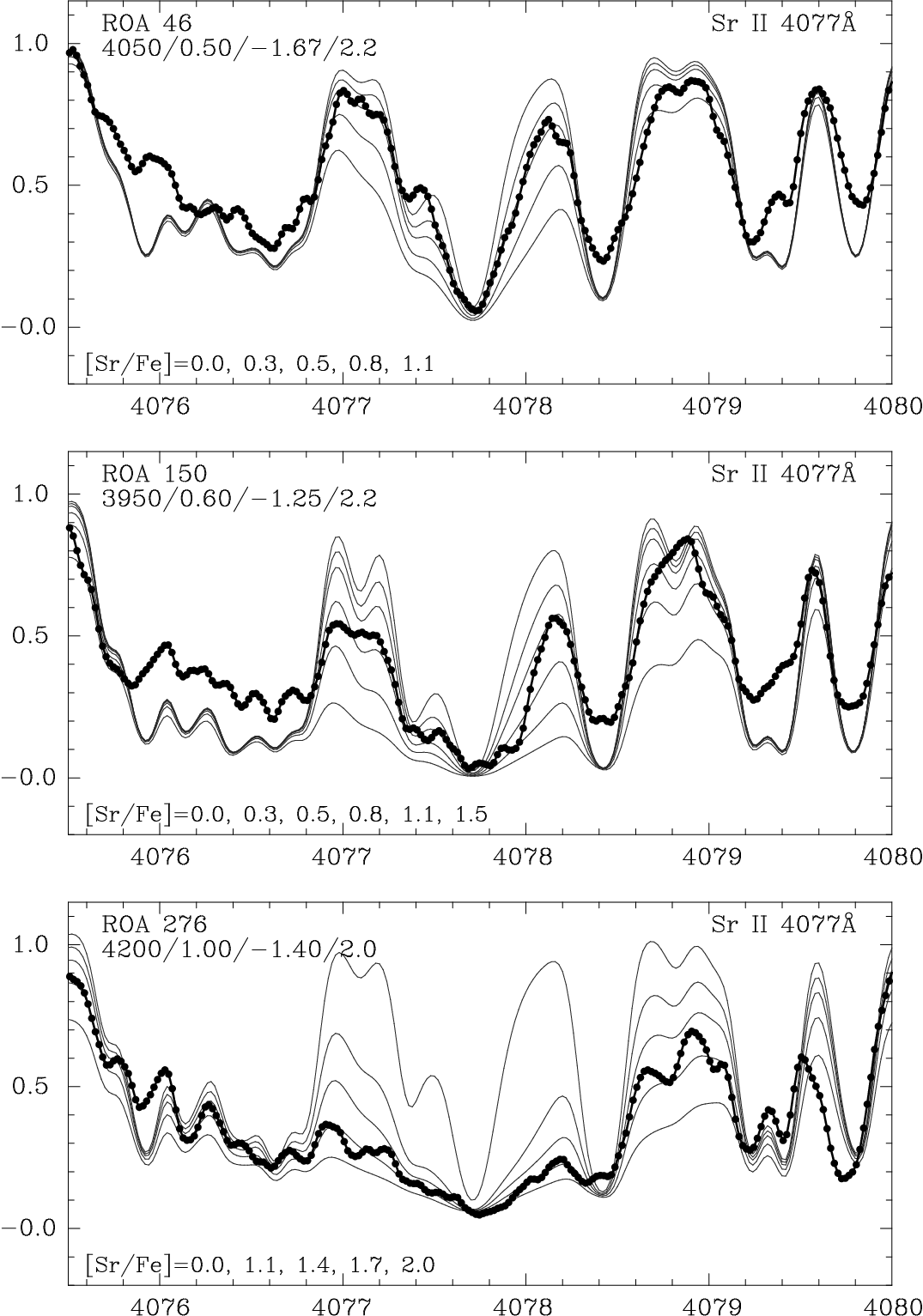}
\caption{ High resolution observed {\sc mike} spectra of ROA 46, 150
  and 276 (black, dotted line) showing the synthetic spectra fit (grey
  lines) to the Sr 4077{\AA} line.  The synthetic spectra have stellar
  parameters: $T_{\rm eff} = 4050$, ${\rm log}g = 0.50$, ${\rm [Fe/H]} =
  -1.67$ and ${\rm v}_{t} = 2.2$ for ROA 46; $T_{\rm eff} = 3950$,
  ${\rm log}g = 0.60$, ${\rm [Fe/H]} = -1.25$ and ${\rm v}_{t} = 2.2$ for
  ROA 150; and $T_{\rm eff} = 4200$, ${\rm log}g = 1.00$, ${\rm [Fe/H]} =
  -1.40$ and ${\rm v}_{t} = 2.0$ for ROA 276.  \label{psr} }
\end{center}
\end{figure}

\begin{figure}
\begin{center}
\includegraphics[width=12.0cm,angle=0]{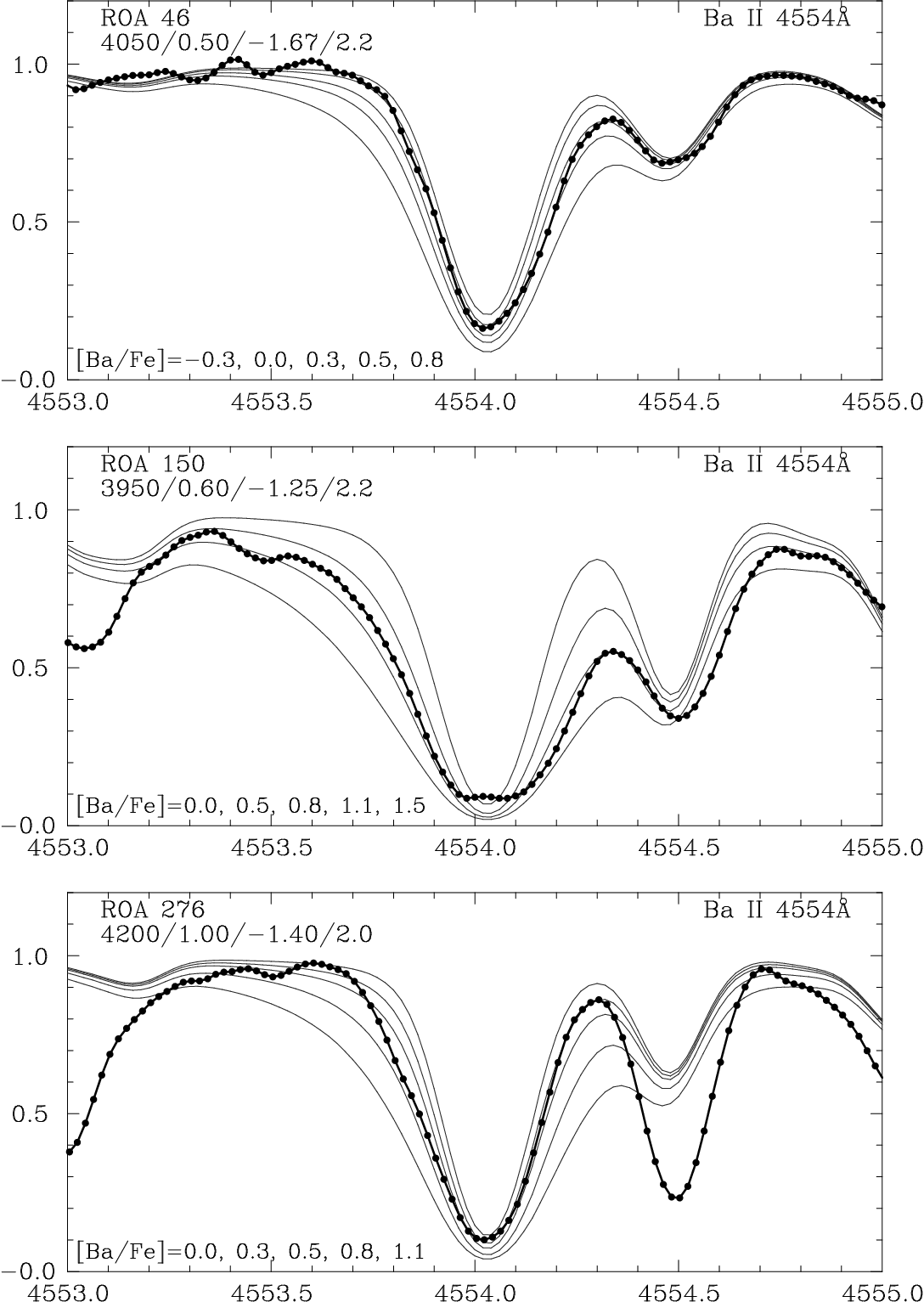}
\caption{ High resolution {\sc mike} spectra of ROA 46, 150 and 276
  (black, dotted line) showing the synthetic spectra fit (grey lines)
  to the Ba 4554{\AA} line.  The synthetic spectra have stellar
  parameters: $T_{\rm eff} = 4050$, ${\rm log}g = 0.50$, ${\rm [Fe/H]}
  = -1.67$ and ${\rm v}_{t} = 2.2$ for ROA 46; $T_{\rm eff} = 3950$,
  ${\rm log}g = 0.60$, ${\rm [Fe/H]} = -1.25$ and ${\rm v}_{t} = 2.2$
  for ROA 150; and $T_{\rm eff} = 4200$, ${\rm log}g = 1.00$, ${\rm
    [Fe/H]} = -1.40$ and ${\rm v}_{t} = 2.0$ for ROA 276.
  \label{pba} }
\end{center}
\end{figure}

Figure \ref{psr} shows the Sr 4077{\AA} line for ROA 40 (panel a), 150
(panel b) and 276 (panel c).  The results for the Ba 4554{\AA} line
are shown for all three stars in Figure \ref{pba} (for ROA 150, one
will notice a flat-bottom in the line profile for the Ba 4554{\AA}
line.  The exact cause of this is unclear, but it is suspected to be
due to saturation effects).  For comparison, in both figures a
synthetic spectrum with stellar parameters $T_{\rm eff} = 4200K$,
${\rm log}g = 1.0$, ${\rm [Fe/H]} = -1.40$ and ${\rm v}_{t} =
2.0kms^{-1}$ have been plotted for a range of [Sr/Fe] and [Ba/Fe]
values against the observed ROA 276 one.  These were chosen as they
gave the best fit to the observed spectra.  Table \ref{tbl7} lists the
[Sr/Fe] and [Ba/Fe] results for each of the three stars as well as the
[Ba/Sr] abundance.  It is clear from the {\sc mike} high resolution
data that ROA 276 has an unusual [Ba/Sr] ratio, and that further study
of this star is warranted.

\section{Discussion} \label{disc}

The abundances found for the RGB stars here can not only provide
information about enrichment sources for these stars, but also by
comparing the abundances with those found for the main sequence and
turnoff, further constraints can be derived.  Comparing the MSTO and
RGB abundance patterns may also show how these enrichments originated
--- that is, whether they are primordial, evolutionary or the material
was accreted onto the surface layers of the stars.

As a star ascends the RGB, the convective envelope deepens
considerably and material that was processed via the CN cycle in the
interior of the star is possibly mixed to the surface layers (see
\citealt{sm79}).  This would result in less carbon and more nitrogen
when compared with earlier evolutionary stages, with the overall C+N
abundance remaining constant.  In this instance the abundances of the
s-process elements are not changed.  The level of mixing, if any, on
the RGB is, however, uncertain.  In some globular clusters there is
evidence of mixing as stars ascend the RGB as the carbon and nitrogen
abundances are different from the MS level compared with those at the
tip of the RGB \citep{lan86, bcs02}.  An example of such a cluster is
M13 (\citealt{sb06}, and references therein).  It shows chemical
inhomogeneities for elements C through to Al, and is a result of a
primordial abundance spread, coupled with mixing on ascent of the RGB.
On the other hand, other globular clusters, such as {\tuc} show no
difference in the abundance patterns of carbon and nitrogen between
the MS and RGB \citep{can98, bri04, dac04}.

A competing factor to the mixing process is the enhancements of
carbon, nitrogen and s-process elements in stars on the MSTO of
{\wcen}.  These enhancements arose either from primordial enrichment
(the enhanced material is uniform throughout the star's interior) or
accretion events (the enhanced material is in the surface layers
only).  In the case of the enhanced material being accreted, as a star
ascends the RGB, the convective envelope deepens and the enhanced
material is mixed with unenhanced material.  This results in a
dilution of the enhanced material which may include carbon, nitrogen
and s-process elements.  Alternatively, if the material is uniform
throughout the star, the same abundances for the MSTO and RGB stars
should be observed.  That said, the evolutionary mixing, described
above, adds a further complication to the picture and must be taken
into consideration when attempting to decipher the differences between
the MSTO and RGB.

Another type of mixing to take into consideration is thermohaline
mixing \citep{krt80, cha07}.  This process involves mixing of material
that has been accreted when the mean molecular weight of the stellar
gas increases towards the surface of the star.  The gas is displaced
downwards and compressed, making it hotter than its surroundings.
This in turn makes it lose more heat, become denser and continue to
sink.  This then leads to mixing on thermal timescales.  \citet{sg08}
estimate thermohaline processes mixing the accreted material with
16---88\% of the pristine gas.  This, however, was for a model of a
star with $Z = 10^{-4}$ (${\rm [Fe/H]} = -2.3$) whereas most of the
enhanced stars are more metal-rich than this by at least one dex.
Therefore the difference in mean molecular weight is smaller, and
thermohaline mixing to be less efficient.  Several other studies have
found the thermohaline process to be less efficient that first
thought.  In a comparison between turnoff stars and red giants,
\citet{aok08} showed that the distribution of [C/Fe] was different
between the two sets, suggesting significant mixing only occurs at the
first dredge-up.  \citet{dp08} reach a similar conclusion using the
data from \citet{luc06}, where they found the [C/H] reduced by
$\sim0.4 {\rm dex}$ resulting from the first dredge up. 

It is interesting to estimate the dilution factor --- the amount surface
pollution is diluted as the convective envelope deepens on ascent of
the RGB.  This factor depends on the metallicity and mass of the star
\citep{yos81} as these influence the mass fraction of the convective
envelope on the MS and RGB. The convective envelope mass faction
increases for increasing metallicity, and decreasing stellar mass.

The dilution factor can then be calculated, given by the following
equation:

\begin{equation} \label{eq2}
\mbox{[X/Fe]}_{\mbox{\small{\textsc{rgb}}}} = log[1+(10^{\mbox{[X/Fe]}_{\mbox{\small{\textsc{ms}}}}}-1) \frac{m_{\mbox{\small{\textsc{ms}}}}}{m_{\mbox{\small{\textsc{rgb}}}}} ] 
\end{equation}

\noindent where ${\rm [X/Fe]}_{\rm {\textsc{ms}}}$ is the abundance of
element X in the surface layers before dilution, ${\rm
  [X/Fe]}_{\rm{\textsc{rgb}}}$ is the abundance of element X on the
RGB after dilution, and the mass fractions of the convective envelope
on the MS and RGB are represented by ${\rm m}_{\rm {\textsc{ms}}}$ and
${\rm m}_{\rm {\textsc{rgb}}}$, respectively.  The interior of the MS
star is assumed to have a solar abundance ratio of [X/Fe].  Using the
convective envelope mass fractions of ${\rm m}_{\rm {\textsc MS}} =
2.256 \times 10^{-2}$ and ${\rm m}_{\rm {\textsc RGB}} = 0.6837$ for
an $0.8{\rm M}_{\odot}$ star with initial metallicity ${\rm [Fe/H]} =
-0.75$ and enhanced $\alpha-{\rm elements}$ (D.\ VandenBerg 2009,
private communication), the depletion factor is found to be $\sim 30$.
Using envelope mass fractions for lower metallicity stars, larger mass
stars, or lower $\alpha-{\rm element}$ enhancement all result in
greater depletion values.  The models of Girardi (2009, priv
communication) give similar results.  This dilution factor indicates
that any surface contamination would be diluted to a large degree,
although still measurable on the RGB.  The inclusion of thermohaline
mixing would only serve to increase the abundance one would find on
the MS using the RGB abundance ratios.

There are several possible enrichment sources for the enhancements.
One possibility is low mass ($\lesssim 3{\rm M}_{\odot}$) AGB stars,
the products of which are carbon and s-process elements \citep{gal98}.
Intermediate mass (3$\lesssim$ ${\rm M}_{\odot} \lesssim 8$) AGB stars
may also play a role, producing nitrogen and perhaps a small amount of
s-process elements (depending on the size of the $^{13}$C pocket)
\citep{vdm02}.  Massive stars may also contribute to the enrichment, in
particular massive rotating objects.  These produce excess helium and
nitrogen in their stellar winds without carbon and oxygen
overabundances \citep{mm06}.  These stars may be able to account for
the abundances found on the RGB.

Those objects with near solar abundance ratios of carbon and
supersolar s-process element enrichment have possibly been enriched by
low mass AGB stars.  Yet objects with enrichment in nitrogen are more
consistent with enhancement from intermediate mass AGB stars. As many
of the nitrogen enhanced objects also show enrichment in s-process
elements several different sources may be required to explain the
observed abundance patterns.  In these cases, rotating, massive stars
may account for the N abundance patterns, while low mass AGB stars
could be the source of the s-process enhancements.  As yet s-process
element abundance yields for rotating, massive stars have not been
published and therefore definitive conclusions regarding the
enhancement sources of enriched nitrogen and strontium RGB stars
cannot be made.

\subsection{Comparison of abundances on the MSTO and RGB}

The RGB stars studied here do not have carbon abundances greater than
${\rm [C/Fe]} = 0.2$, except for the CH star ROA 279 (${\rm [C/Fe]} =
0.6$), while few MSTO stars show depletions and several show
enhancements (${\rm [C/Fe]}_{{\rm MSTO}} = 0.5-1.0$).  Consequently,
the low [C/Fe] values on the RGB may be due to the evolutionary
effects during ascent of the RGB involving processing of C into N.  It
should be kept in mind, however, that our ability to measure the C
abundance for ${\rm [C/Fe]} < 0.5$ for the metal-poor and intermediate
metallicity MSTO stars is not high; therefore detailed comparisons,
for example as a function of [Fe/H], are not easily made.

\begin{figure}
\begin{center}
\includegraphics[width=7.0cm,angle=0]{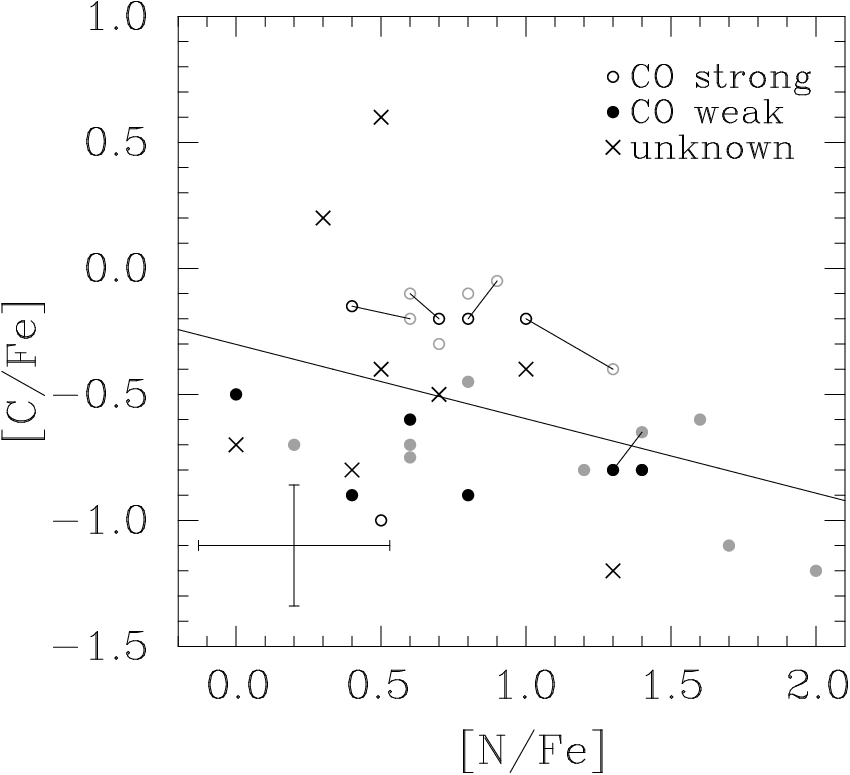}
\caption{ 
Carbon abundance plotted as a function of nitrogen abundance for the
R600 (grey) and  (black) resolution samples. The open circles
represent the CO strong stars, and closed circles represent the CO
weak stars.  The stars for which there is no information regarding the
CO nature are represented by crosses. Abundances of stars in common
between the R600 and R1200 resolution analysis are joined by solid
lines.  The solid line represents the least squares fit to the data.
\label{pcnb} }
\end{center}
\end{figure}

Apart from the CH star ROA 279, many RGB stars are carbon depleted as
would be expected from convective mixing during ascent of the RGB.
However, inspection of Fig. \ref{pcnb} shows a significant number of
stars with ${\rm [C/Fe]} \sim 0.0$, which were previously found in
\citetalias{nd95b} (the CO-strong objects of \citet{per80}), and in
comparison with other globular clusters these objects are unique.
\citetalias{nd95b} provided three possible explanations for the carbon
abundances of these stars: (1) They are only found in {\wcen} due to
the larger number of stars sampled in this cluster; (2) carbon
depletion as a star ascends the RGB does not occur for all stars; and
(3) depletion does occur in all stars, and the initial carbon
abundance of those objects with near solar [C/Fe] was considerably
higher than for the bulk of the population before its ascent of the
RGB.

The carbon results found on the main sequence and turnoff are
consistent with the third possibility. The MSTO objects were found to
have enhancements of ${\rm [C/Fe]} \geq 0.5$ for $\sim 10\%$ of stars
(see Table 3, \citet{sta07}).  \citetalias{nd95b} suggest, in their
consideration of the unbiased samples of \citet{per80} and
\citet{cs73} that within 1 mag of the tip of the RGB in $\omega$ Cen
$5-15\%$ of the stars are CO-strong.  As we do not find stars with
carbon enhancements as great as those found at the MSTO on the upper
RGB (except for the four CH stars), mixing to some degree must have
occurred during the ascent of the RGB and resulted in objects with
carbon abundances higher than found in the bulk of the main RGB
population (i.e. ${\rm [C/Fe]} > \sim-0.2$).  \citetalias{nd95b}
reported depletion of carbon at the RGB tip in the range $\Delta{\rm
  [C/Fe]} = 0.3 - 1.0$ dex, supporting the above result.  Therefore,
the carbon rich objects on the MSTO could be the evolutionary
precursors to the CO-strong stars on the RGB.

On inspection of Figure \ref{pcnb} it can be seen that those stars
with the greatest depletions of carbon have the greatest enhancements
of nitrogen.  Those with near solar carbon abundance ratios have more
modest nitrogen enhancements.  This may be accounted for if the {\it
  amount} of evolutionary mixing a star goes though is not the same
for every star for a given mass and metallicity.  Main sequence
objects, on ascent of the RGB, may have processed little C into N
resulting in RGB stars with near solar [C/Fe], and moderate N
enhancements. Alternatively, they may have processed a more
substantial amount of carbon resulting in objects with low carbon
abundances (${\rm [C/Fe]} \sim -0.8$, and large nitrogen
enhancements.  Nitrogen enhancements on the RGB show high ${\rm
  [N/Fe]} = 0.5 - 1.5$.  This may be due to two sources. The first is
primordial enrichment, the second CN cycling and mixing as the star
ascends the RGB.

\subsubsection{Carbon rich stars}

Two main sequence C-enhanced stars, S7007334 (${\rm [Fe/H]} = -1.84$)
and S9005309 (${\rm [Fe/H]} = -1.88$), were identified in
\citet{sta07} that may be the counterparts to the RGB CH stars in the
cluster.  They show enhancements in [C/Fe] greater than those found
for the CH star ROA 279 (${\rm [Fe/H]} = -1.69$) by 0.6~dex.  These MS stars
show no nitrogen or s-process enhancements that CH stars usually
exhibit, although this may be due to the low sensitivity of our
technique to these features at MS metallicities and temperatures.

On the subgiant branch (SGB), the star S8001811 \citep{sta07} is
worthy of mention.  Its carbon, nitrogen and strontium abundance
pattern (${\rm [C/Fe]} = 0.7$, ${\rm [N/Fe]} = 1.2$, ${\rm [Sr/Fe]} =
1.4$) is similar to that seen in the CH star ROA 279.  This SGB object
may be a precursor of the RGB CH stars in {\wcen}, and a counterpart
of the field ``subgiant CH'' stars of \citet{lb91}.

\subsubsection{Nitrogen and s-process enhanced stars}

There exists a number of metal-rich stars (${\rm [Fe/H]} > -1.1$) on
the MSTO that show enhancements in both nitrogen and s-process
elements, together with low carbon, (${\rm [C/Fe]} \sim -0.3$, ${\rm
  [N/Fe]} \sim 1.8$, ${\rm [Sr/Fe]} \sim 1.0$) \citep{sta07}.  On the
RGB, ROA 480 shows similar abundance patterns and is of similar
metallicity with ${\rm [Fe/H]} = -0.95$.  With ${\rm [C/Fe]} =
-0.6$, its carbon abundance is depleted by a larger amount than that
of the MSTO group.  This difference may be due to the carbon having
been processed into nitrogen as ROA 480 ascended the RGB.  ROA 480
(with ${\rm [N/Fe]} = 0.6$, ${\rm [Sr/Fe]} = 0.6$) has lower nitrogen
and strontium than found for the MSTO group.  These abundance patterns
with low carbon, and enhanced nitrogen and strontium are also seen at
lower metallicities.  For example, ROA 84 (${\rm [Fe/H]} = -1.35$)
and a star on the main sequence, S5004811 (${\rm [Fe/H]} = -1.40$),
share similar abundances in carbon and nitrogen.  Both have carbon
abundances of ${\rm [C/Fe]} = -0.1$, and high nitrogen (${\rm  [N/Fe]}_{\rm RGB} = 0.8$ and ${\rm [N/Fe]}_{\rm MSTO} = 1.7$).
These abundances agree within the errors of measurement.  The
s-process enhancements have somewhat different values, with ${\rm
  [Sr/Fe]}_{\rm RGB} = 0.5$ and ${\rm [Sr/Fe]}_{\rm MSTO} = 1.3$.
\citetalias{nd95b} give high resolution abundances for both ROA 84 and
480 which are consistent with these results.  For ROA 84 ${\rm [C/Fe]}
= 0.05$, ${\rm [N/Fe]} = 0.35$ and ${\rm [s/Fe]} = 0.45$ (where [s/Fe]
is the average values of [Y/Fe] and [Zr/Fe]), while for ROA 480 ${\rm
  [C/Fe]} = -0.65$, ${\rm [N/Fe]} = 0.80$ and ${\rm [s/Fe]} = 0.38$.

The upper levels of nitrogen enhancements on the MSTO and RGB are
similar, with ${\rm [N/Fe]} \sim 1.8$.  This may indicate that enhanced N
abundances are driven more by a phenomenon that is induced in the
cluster as a whole. The upper limit on nitrogen enhancements may also
indicate that mixing on ascent of the RGB does not occur to the same
extent for all stars.  If it did, then higher [N/Fe] abundances should
be measured for objects on the RGB than for those on the MSTO as an
increase in nitrogen occurs in the CN cycle (although given the errors
of measurement a difference in the measured abundance between the MSTO
and RGB may not be reliable).  However, this makes the assumption that
the enriched material on the MSTO is uniform throughout the star.  In
the case where MSTO objects have accreted material onto their surface
layers, on ascent of the RGB the convective envelope deepens and
dilution of the N enhanced material occurs.  Coupled with this is the
increase in [N/Fe] due to internal processing and mixing.  As the two
scenarios (primordial or  accretion event) cannot be differentiated due
to the complication of mixing on the ascent of the RGB, abundances of
elements not effected by this process need to be investigated and
compared at the different evolutionary stages. An example of such an
element is Sr, and a comparison of the Sr abundances found for MSTO
and RGB stars is described in the following section.

\subsubsection{S2015448 \& ROA 276} \label{msrgb}

As noted in \S\ref{results}, ROA 276 stood out against all other
members of our RGB sample.  For our initial adopted metallicity of
${\rm [Fe/H]} = -0.57$, we obtained ${\rm [Ba/Sr]} = -1.2$ from our
high-resolution {\sc mike} spectra (see Table \ref{tbl6}). (We also
found in \S\ref{roa276} that this value is quite insensitive to
uncertainties in atmospheric parameters). In contrast, the [Ba/Fe]
value was found to be equal to or greater than [Sr/Fe] for every other
star analyzed.

This low [Ba/Sr] value for ROA 276 is similar to that found for the
main sequence star S2015448 \citep{sta06b}, which has ${\rm [Sr/Fe]} =
1.6$, ${\rm [Ba/Fe]} < 0.6$ and ${\rm [Ba/Sr]} < -1.0$, as may be
seen in Table \ref{tbl7}.  While the relative frequencies of such
large Sr enhancements on the MSTO and RGB are not well determined, and
must await further work, the existence of such objects on both the MS
and RGB places an important constraint on the origin of the
enhancement.  Specifically, unless accretion provides a large fraction
of the mass of the observed star, a primordial origin of the
enhancement is the more likely explanation for the anomaly.  That is
to say, if the Sr enhancement in S2015448 is confined only to its
small convective outer region following accretion of only a small
fraction of the star's mass, convective dilution, as described by
equation 6.2, will lead to a reduced relative abundance, ${\rm [Sr/Fe]} = 0.35$,
when it reaches the RGB.  Conversely (and perhaps more to the point),
given the same assumptions, one would conclude that ROA 276 would have
had the enormous value, ${\rm [Sr/Fe]} \approx 2.3$, when it was on the main
sequence if it was surface contaminated rather than primordially
enhanced.  While not outside the realms of possibility, this is
unlikely.

This argument can be applied to many of the stars on the RGB with
s-process enhancements (the carbon and nitrogen abundances are
complicated by evolutionary mixing abundance changes) similar to those
found at the MSTO (to within errors of measurement).  The abundance
enhancements seen in these stars are unlikely to have been surface
accretion phenomenon unless the mass accreted is very
substantial. For example, ROA 279 with ${\rm [C/Fe]} = 0.6$ on the RGB
would have a carbon abundance of ${\rm [C/Fe]} > 2$ on the MS.  No
stars with such a high carbon abundance ratio were found on the MS.

\section{Conclusions}

Abundances of C, N and the s-process elements Sr and Ba were
determined for a biased sample of 33 RGB stars in {\wcen}.  Almost all
objects show depletion of carbon, and solar or enhanced nitrogen.  The
abundances of Sr and Ba show enhancement as well.  One of the known CH
stars in {\wcen} has been analyzed for the first time for carbon and
nitrogen, resulting in ${\rm [C/Fe]} = 0.6$, ${\rm [N/Fe]} = 0.5$ and
${\rm [Sr/Fe]} = {\rm [Ba/Fe]} = 0.6$.  This star is less enhanced in
carbon compared with other CH stars in the cluster, but still has
considerable enhancement in carbon compared with other RGB stars.

The levels of N enhancement on both the MS and RGB reach similar
relative abundances, ${\rm [N/Fe]} \sim 1.8$.  This may indicate that
mixing on the RGB occurs to different extents for individual stars.

A RGB star with high enhancement of the light s-process element Sr was
found, but with little enhancement in Ba.  This star is similar to the
strongly Sr-enhanced MS object, S2015448 \citep{sta06b} from which the
conclusion is reached that the Sr enhancement is likely to be
primordial in origin, rather than the result of some accretion event.
\\

{\bf Acknowledgments}

We wish to thank Don VandenBerg and Leo Girardi for providing
unpublished details from their models.  Australian access to the
Magellan Telescopes was supported through the Major National Research
Facilities II program of the Australian Government.  We thank the
referee for their comments which led to improvements in the
manuscript. LMS wishes to thank K. Ward for advice on statistical
calculations.

{\it Facilities:} {ANU:SSO2.3m(DBS); Magellan:Clay(MIKE)}


\clearpage
\begin{table}[!hb]
\begin{center}
\caption[List of RGB objects]{List of RGB objects with observing dates and grating used. \label{tbl1}}
\begin{tabular}{lll}
\tableline\tableline
Grating  & Date & Star ID  \\
(1)              & (2)            & (3)      \\
\tableline
600 I            & May 2005       & ROA: 24, 40, 43, 46, 53, 58, 84, 132, 150, 159 \\
                 &                & ROA: 162, 171, 179, 219, 248, 252, 253, 371 \\
                 &                & \\
1200 I           & Feb/Jun  2002  & ROA: 40, 42, 43, 46, 48, 55, 65, 74, 84, 94 139 \\
                 &                & ROA: 159, 179, 201, 213, 219, 253, 276, 279, 300\\
                 &                & ROA: 324, 336, 357, 425, 447, 477 , 480, 517, \\
                 &                & OC: 140419, 222068,  230189, 242056, 250606, 263340\\
                 &                & OC: 305654, 312058, 321293, 618774, 801639\\
\tableline
\end{tabular}
\end{center}
\end{table}

\clearpage
\begin{deluxetable}{lccccccc}
\tabletypesize{\scriptsize}
\tablecolumns{8}
\tablewidth{0pt}
\tablecaption{Stellar Parameters for RGB stars in $\omega$ Cen \label{tbl2}}
\tablehead{
\colhead{ID}&\colhead{V\tablenotemark{1}}&\colhead{B--V\tablenotemark{2}}&\colhead{Teff}&\colhead{log$g$}&\colhead{[Fe/H]}&\colhead{V$_{t}$}&\colhead{source\tablenotemark{3}}
}
\startdata
ROA 40         &  11.37   & 1.48    & 4200  & 0.5    & --1.69 & 2.3   & 1   \\ 
ROA 42         &  11.64   & 1.49    & 4150  & 0.5    & --1.69 & 2.0   & 1   \\
ROA 43         &  11.62   & 1.62    & 3950  & 0.4    & --1.47 & 2.1   & 1   \\
ROA 46         &  11.54   & 1.55    & 4050  & 0.5    & --1.67 & 2.2   & 1   \\ 
ROA 48         &  11.51   & 1.58    & 4050  & 0.5    & --1.76 & 2.5   & 1   \\
ROA 53         &  11.58   & 1.64    & 3950  & 0.4    & --1.67 & 2.3   & 1   \\
ROA 58         &  11.67   & 1.43    & 4200  & 0.6    & --1.73 & 2.2   & 1   \\
ROA 65         &  11.61   & 1.50    & 4050  & 0.6    & --1.72 & 2.1   & 1   \\
ROA 74         &  11.78   & 1.37    & 4250  & 0.7    & --1.80 & 2.2   & 1   \\
ROA 84         &  11.87   & 1.66    & 3900  & 0.5    & --1.36 & 1.9   & 1   \\
ROA 94         &  11.80   & 1.39    & 4200  & 0.7    & --1.78 & 2.1   & 1   \\
ROA 132        &  12.05   & 0.87    & 3900  & 0.3    & --1.37 & 2.2   & 1   \\
ROA 139        &  11.98   & 1.45    & 4150  & 0.8    & --1.46 & 1.6   & 1   \\
ROA 150        &  12.00   & 1.69    & 3950  & 0.6    & --1.25 & 2.2   & 1   \\
ROA 159        &  12.04   & 1.34    & 4300  & 0.9    & --1.72 & 2.0   & 1   \\
ROA 162        &  12.14   & \nodata & 3950  & 0.7    & --1.10 & 2.1   & 1   \\
ROA 171        &  12.07   & 1.47    & 4100  & 0.7    & --1.43 & 1.9   & 1   \\
ROA 179        &  12.21   & 1.65    & 3850  & 0.5    & --1.10 & 1.5   & 1   \\
ROA 213        &  12.22   & 1.12    & 4500  & 1.1    & --1.83 & 1.9   & 1   \\
ROA 219        &  12.20   & \nodata & 4000  & 0.7    & --1.10 & 2.2   & 1   \\
ROA 248        &  12.43   & 1.70    & 3850  & 0.6    & --0.78 & 1.6   & 1   \\
ROA 252        &  12.30   & 1.20    & 4400  & 1.1    & --1.74 & 2.0   & 1   \\
ROA 253        &  12.34   & 1.37    & 4300  & 1.0    & --1.39 & 1.9   & 1   \\
ROA 276        &  12.37   & 1.38    & 4000  & 0.7    & --0.57 & 2.0   & 3,5 \\
ROA 279        &  12.32   & \nodata & 4350  & 1.1    & --1.69 & 2.0   & 1   \\
ROA 300        &  12.71   & 1.68    & 3900  & 0.7    & --0.77 & 2.0   & 2,3 \\
ROA 324        &  12.59   & 1.54    & 4000  & 0.7    & --0.39 & 1.9   & 3,5 \\
ROA 357        &  12.69   & 1.43    & 4000  & 0.8    & --0.85 & 1.8   & 1   \\
ROA 371        &  12.68   & 1.61    & 4000  & 0.9    & --0.79 & 1.6   & 1   \\
ROA 425        &  12.73   & \nodata & 3650  & 0.6    & --0.32 & 2.0   & 2,3 \\
ROA 447        &  12.80   & 1.64    & 3700  & 0.7    & --0.18 & 2.0   & 2,3 \\
ROA 480        &  12.98   & 1.25    & 4350  & 1.3    & --0.95 & 1.8   & 1   \\
OC 140419      &  13.49   & 1.22    & 4200  & 1.5    & --0.65 & 2.0   & 4   \\
OC 263340      &  13.62   & 1.13    & 4400  & 1.9    & --0.63 & 1.2   & 4   \\
OC 305654      &  13.38   & 1.29    & 4200  & 1.6    & --0.65 & 1.4   & 4   \\
OC 321293      &  13.69   & 1.15    & 4300  & 1.6    & --0.71 & 1.4   & 4   \\
\enddata				       	    
\tablenotetext{1}{NFM96 and references therein; \citet{pan03}}
\tablenotetext{2}{\citet{van00}, \citet{pan03}}
\tablenotetext{3}{Source: ND95 (1); PFCAM (2); NFM96 (3); \citet{pan03} (4); This study (5) }	    
\end{deluxetable}

\clearpage
\begin{deluxetable}{lrrrrrrrrrrrrr}
\tabletypesize{\footnotesize}
\tablecolumns{14}
\tablewidth{0pt}
\tablecaption{Abundances for carbon, nitrogen, strontium and barium, and the individual errors for the R600 data. \label{tbl3}}
\tablehead{
\colhead{ID} & \colhead{[Fe/H]} &\colhead{} &\multicolumn{2}{c}{Carbon} &\colhead{} & \multicolumn{2}{c}{Nitrogen} &\colhead{} & \multicolumn{2}{c}{Strontium} &\colhead{} & \multicolumn{2}{c}{Barium}\\
\colhead{}   & \colhead{}       &\colhead{} &\colhead{ab\tablenotemark{1}} & \colhead{e\tablenotemark{1}} &\colhead{} & \colhead{ab\tablenotemark{1}} & \colhead{e\tablenotemark{1}}   &\colhead{} & \colhead{ab\tablenotemark{1}} & \colhead{e\tablenotemark{1}}    &\colhead{} & \colhead{ab\tablenotemark{1}} &\colhead{e\tablenotemark{1}}
}
\startdata
ROA 40     & --1.69 & & --0.20 & 0.24 &  & 0.6  & 0.31 &  & 0.3   & 0.36 &  & 0.5   & 0.32     \\
ROA 43     & --1.47 & & --0.30 & 0.24 &  & 1.2  & 0.38 &  & 0.4   & 0.36 &  & 0.5   & 0.18     \\
ROA 46     & --1.67 & & --0.70 & 0.24 &  & 0.5  & 0.38 &  & 0.0   & 0.44 &  & 0.3   & 0.32     \\
ROA 53     & --1.67 & & --0.40 & 0.29 &  & 0.7  & 0.38 &  & --0.1 & 0.44 &  & 0.4   & 0.32     \\
ROA 58     & --1.73 & & --0.80 & 0.29 &  & 0.6  & 0.31 &  & --0.1 & 0.36 &  & 0.1   & 0.32     \\
ROA 84     & --1.36 & & --0.10 & 0.24 &  & 0.9  & 0.35 &  & 0.7   & 0.28 &  & 0.9   & 0.18     \\
ROA 132    & --1.37 & & --0.50 & 0.29 &  & 0.9  & 0.35 &  & 0.0   & 0.31 &  & 0.4   & 0.22     \\
ROA 150    & --1.25 & & --1.00 & 0.24 &  & 1.5  & 0.38 &  & 0.6   & 0.28 &  & 0.7   & 0.18     \\
ROA 159    & --1.72 & & --0.50 & 0.24 &  & 0.2  & 0.38 &  & --0.2 & 0.44 &  & --0.2 & 0.41     \\
ROA 162    & --1.10 & & --0.60 & 0.24 &  & 1.5  & 0.38 &  & 0.3   & 0.36 &  & 0.8   & 0.22     \\
ROA 171    & --1.43 & & --0.20 & 0.29 &  & 0.9  & 0.31 &  & 0.7   & 0.28 &  & 0.6   & 0.22     \\
ROA 179    & --1.10 & &   0.10 & 0.24 &  & \nodata  & 0.35 &  & 0.5   & 0.28 &  & 0.8   & 0.18     \\
ROA 219    & --1.10 & &   0.00 & 0.24 &  & 0.5  & 0.31 &  & 0.4   & 0.28 &  & 0.7   & 0.18     \\
ROA 248    & --0.78 & & --1.00 & 0.29 &  & 1.8  & 0.38 &  & 0.6   & 0.31 &  & 0.6   & 0.18     \\
ROA 253    & --1.39 & & --0.70 & 0.24 &  & 1.4  & 0.31 &  & 0.4   & 0.36 &  & 0.8   & 0.32     \\
ROA 371    & --0.79 & & --0.90 & 0.29 &  & 1.4  & 0.35 &  & 0.4   & 0.31 &  & 0.9   & 0.27     \\
\enddata	
\tablenotetext{1}{ab=[X/Fe], e=$\Delta$[X/Fe]}			       	    
\end{deluxetable}

\clearpage
\begin{deluxetable}{lrrrrrrrrrrrrr}
\tabletypesize{\footnotesize}
\tablecolumns{14}
\tablewidth{0pt}
\tablecaption{Abundances for carbon, nitrogen, strontium and barium, and the individual errors for the R1200 data. \label{tbl4}}
\tablehead{
\colhead{ID} & \colhead{[Fe/H]} &\colhead{} &\multicolumn{2}{c}{Carbon} &\colhead{} & \multicolumn{2}{c}{Nitrogen} &\colhead{} & \multicolumn{2}{c}{Strontium} &\colhead{} & \multicolumn{2}{c}{Barium}\\
\colhead{}   & \colhead{}       &\colhead{} &\colhead{ab\tablenotemark{1}} & \colhead{e\tablenotemark{1}} &\colhead{} & \colhead{ab\tablenotemark{1}} & \colhead{e\tablenotemark{1}}   &\colhead{} & \colhead{ab\tablenotemark{1}} & \colhead{e\tablenotemark{1}}    &\colhead{} & \colhead{ab\tablenotemark{1}} &\colhead{e\tablenotemark{1}}
}
\startdata
ROA 40	  & --1.69 & & --0.10   & 0.16    & & 0.4     & 0.22    & & --0.1     & 0.34    & & 0.3     & 0.28    \\
ROA 43 	  & --1.47 & & --0.40   & 0.21    & & 1.0     & 0.26    & & 0.4      & 0.37    & & 0.8     & 0.31    \\
ROA 46	  & --1.67 & & --0.90   & 0.25    & & \nodata & 0.43    & & \nodata  & \nodata & & 0.0     & 0.44    \\
ROA 48	  & --1.76 & & --0.80   & 0.18    & & 0.4     & 0.43    & & --0.2     & 0.40    & & --0.2    & 0.44    \\
ROA 65	  & --1.72 & & --0.60   & 0.18    & & \nodata & \nodata & & \nodata  & \nodata & & \nodata & \nodata \\
ROA 74	  & --1.80 & & --0.80   & 0.18    & & 0.8     & 0.43    & & 0.0      & 0.34    & & 0.0     & 0.36    \\
ROA 84	  & --1.36 & & --0.10   & 0.25    & & 0.8     & 0.30    & & 0.5      & 0.37    & & 0.5     & 0.36    \\
ROA 94	  & --1.78 & & --0.70   & 0.18    & & 0.0     & 0.43    & & 0.0      & 0.34    & & 0.0     & 0.36    \\
ROA 139	  & --1.46 & & --0.90   & 0.21    & & 0.5     & 0.34    & & 0.0      & 0.34    & & 0.2     & 0.31    \\
ROA 179	  & --1.10 & &  0.00    & 0.25    & & \nodata & 0.34    & & 0.3      & 0.34    & & 0.5     & 0.28    \\
ROA 213	  & --1.83 & & --0.50   & 0.18    & & 0.0     & \nodata & & \nodata  & \nodata & & \nodata & \nodata \\
ROA 219	  & --1.10 & & --0.20   & 0.21    & & 0.7     & 0.30    & & 0.1      & 0.34    & & 0.5     & 0.28    \\
ROA 253	  & --1.39 & & --0.70   & 0.21    & & 1.3     & 0.26    & & 0.3      & 0.34    & & 0.7     & 0.31    \\
ROA 276	  & --0.57 & & --0.80   & 0.21    & & 0.4     & 0.34    & & 0.8      & 0.34    & & 0.2     & 0.28    \\
ROA 279	  & --1.69 & &  0.60    & 0.25    & & 0.5     & 0.30    & & 0.6      & 0.34    & & 0.6     & 0.28    \\
ROA 324	  & --0.39 & & --1.00   & 0.34    & & 1.3     & 0.30    & & 0.0      & 0.34    & & 0.0     & 0.28    \\
ROA 357	  & --0.85 & & --0.80   & 0.25    & & 1.4     & 0.30    & & 0.0      & 0.40    & & 0.6     & 0.28    \\
ROA 425	  & --0.32 & & \nodata  & \nodata & & \nodata & \nodata & & 0.0      & 0.40    & & 0.3     & 0.36    \\
ROA 480	  & --0.95 & & --0.60   & 0.21    & & 0.6     & 0.22    & & 0.6      & 0.34    & & 0.6     & 0.28    \\
OC 140419 & --0.65 & & --0.60   & 0.25    & & 0.7     & 0.26    & & 0.3      & 0.34    & & 0 5     & 0.31    \\
OC 263340 & --0.63 & & --0.40   & 0.21    & & 0.5     & 0.22    & & 0.6      & 0.34    & & 0.7     & 0.28    \\
OC 305654 & --0.65 & & --0.30   & 0.25    & & 1.0     & 0.30    & & 0.0      & 0.34    & & 0.4     & 0.36    \\
OC 321293 & --0.71 & &  0.40    & 0.21    & & 0.3     & \nodata & & 0.1      & 0.34    & & 0.5     & 0.31    \\
\enddata				       	    
\tablenotetext{1}{ab=[X/Fe], e=$\Delta$[X/Fe]}			       	    
\end{deluxetable}

\clearpage

\clearpage
\begin{deluxetable}{lrrrrr}
\tablecolumns{6}
\tablewidth{0pt}
\tablecaption{[Ba/Sr] abundances obtained from Magellan/{\sc mike} data as a function of temperature and metallicity for ROA 276. \label{tbl5}}
\tablehead{
& &  \multicolumn{4}{c}{T$_{eff}$} \\
{[Fe/H]} &   & 3800 & 4000 & 4200 & 4400 
}
\startdata
--0.6               & & --1.40 & --1.10 & --1.45 & --1.40 \\
--1.0               & & --1.20 & --1.15 & --1.40 & --1.30 \\
--1.4               & & --1.10 & --1.30 & --1.20 & --1.35 \\
\enddata	
\end{deluxetable}

\begin{deluxetable}{lrrr}
\tablecolumns{4}
\tablewidth{0pt}
\tablecaption{[Sr/Fe] and [Ba/Fe] abundances using {\sc mike} high resolution spectra for three $\omega$ Cen RGB stars --- ROA 46, 150 and 276. \label{tbl6}}
\tablehead{
Star  &    [Sr/Fe] & [Ba/Fe] & [Ba/Sr] 
}
\startdata
ROA 46        & 0.4 & 0.15 & --0.25 \\
ROA 150       & 0.6 & 0.70 & 0.10   \\
ROA 276       & 1.6 & 0.40 & --1.20 \\
\enddata	
\end{deluxetable}

\begin{table}
\begin{center}
\caption{ Comparisons of abundances obtained from R600 2.3m data for ROA 276 with those of S2015448
  \label{tbl7}}
\vspace*{10pt}
\begin{tabular}{lrcccc}
\tableline\tableline
ID                 & [Fe/H] & [C/Fe] & [N/Fe] & [Sr/Fe] & [Ba/Fe]\\
\tableline
ROA 276	           & --0.57 & --0.80 & 0.4    & 0.8     & 0.2 \\	
S2015448 & --0.74 & --0.50 & $<$0.5 & 1.6     & $<$0.6 \\	
\tableline
\end{tabular}
\end{center}
\end{table}

\clearpage

\clearpage

\clearpage

\clearpage

\clearpage

\clearpage

\clearpage

\clearpage

\clearpage

\clearpage

\clearpage

\clearpage

\end{document}